\documentclass[a4paper,11pt]{article}
\pdfoutput=1
\usepackage{enumerate}
\usepackage{graphicx}
\usepackage{amsmath}
\usepackage{amssymb}
\usepackage{amsfonts}
\usepackage{amsthm}
\usepackage[T1]{fontenc}
\usepackage{float}

\def\aeff{a_{e}}
\def\reff{r_{e}}
\def\geff{g_{e}}
\def\Feff{\mathcal{F}_{e}}
\def\Seff{S_{eff}}
\def\cneff{c_{n,e}}

\def\coneeff{c_{1,e}}
\def\ctwoeff{c_{2,e}}
\def\rhoeff{\rho_{e}}

\def\ep{\varepsilon}
\def\half{\frac{1}{2}}

\def\K{\mathcal{K}}
\def\F{\mathcal{F}}
\def\C{\mathcal{C}}
\def\lr#1{\left(#1\right)}
\def\slr#1{\left[#1\right]}

\def\trl#1{\textrm{Tr}\lr{#1}}

\def\avg#1{\left\langle #1\right\rangle}

\def\CPn{\mathbb C P^n}

\def\RS2{\mathbb R\times S^2_F}

\def \be  {\begin{equation}}
\def \ee  {\end{equation}}
\def \bex  {\begin{equation*}}
\def \eex  {\end{equation*}}
\def \bea {\begin{eqnarray}}
\def \eea {\end{eqnarray}}
\def \bal {\begin{align}}
\def \eal {\end{align}}

\def\no{\nonumber\\}
\def\pd#1#2{\frac{\partial #1}{\partial #2}}
\def\labell#1{\label{#1}}

\def \PRD {{Phys. Rev. D\ }}
\def \JHEP {{JHEP\ }}

\allowdisplaybreaks

\begin{document} 
\begin{titlepage}
\null\vspace{-62pt} \pagestyle{empty}
\begin{center}
\vspace{1truein} {\Large\bfseries
Asymmetric hermitian matrix models and fuzzy field theory}\\
\vskip .1in
{\Large\bfseries
~}\\
\vspace{6pt}
\vskip .1in
{\Large \bfseries  ~}\\
\vskip .1in
{\Large\bfseries ~}\\
{\large\sc Juraj Tekel}\\
\vskip .2in
{\itshape Department of Theoretical Physics and Didactics of Physics,\\ Faculty of Mathematics, Physics and Informatics,\\ Comenius University, Mlynsk\'a Dolina, Bratislava, 842 48, Slovakia}\\

\vskip .1in
\begin{tabular}{r l}
E-mail:
&{\fontfamily{cmtt}\fontsize{11pt}{15pt}\selectfont tekel@fmph.uniba.sk}\\

\end{tabular}

\fontfamily{cmr}\fontsize{11pt}{15pt}\selectfont
\vspace{.8in}
\centerline{\large\bf Abstract}
\end{center}

We analyze two types of hermitian matrix models with asymmetric solutions. 
One type breaks the symmetry explicitly with an asymmetric quartic potential. We give the phase diagram of this model with two different phase transitions between the one cut and two cut solutions.
The second type, describing real scalar field theory on fuzzy spaces, breaks the symmetry spontaneously with multitrace terms. We present two methods to study this model, one direct and one using a connection with the first type of models.
We analyze the model for the fuzzy sphere and obtain a phase diagram with the location of the triple point in a good agreement with the most recent numerical simulations.

\end{titlepage}
\pagestyle{plain} \setcounter{page}{2}

\section{Introduction}

Physics on spaces with spacetime noncommutativity has been around for some time now. First considered in 1930's as a possible way out off the singularities of then forming quantum field theory \cite{snyder}, later reappearing as an effective description of various phenomena \cite{ppsug,witten,qhe2}. The appealing feature of these spaces is a discreet short distance structure, which however does not spoil the symmetries of the space. Such structure is believed to be important in any theory that is ambitious enough to be a theory of quantum gravity \cite{doplicher}.

It was however realized that the nonlocality of the field theories defined on noncommutative spaces comes with unexpected consequences \cite{uvir1,uvir2}. Short and long distance phenomena couple in an intricate way and the usual UV divergences appear also as IR divergences in loop calculations. This is the UV/IR mixing phenomenon, which spoils the renormalizability of the theory and persists also in the commutative limit of the underlying space.

Connected to this phenomenon, noncommutative field theories have been shown to have different symmetry breaking patters from the ordinary theories. Usual commutative field theories are known to have two different phases \cite{comr2}. A disorder phase, where the field oscillates around the zero value, and an uniform order phase, where the field oscillates around the true minimum of the potential. This is quite similar to the spontaneous symmetry breaking in the condensed matter physics and the particle physics. Noncommutative field theories have a third phase, a non-uniform order phase, or a striped phase, where the field does not oscillate around one single value in the whole space \cite{NCphase1}. This phase breaks the translational symmetry of the underlying space and most importantly persist in the commutative limit, being related to the UV/IR mixing.

This third phase has been observed in multiple numerical simulations for the fuzzy sphere \cite{num09,num14}, the fuzzy disc \cite{num_disc}, the fuzzy sphere with a commutative time \cite{num_RSF2} and also for noncompact noncommutative plane \cite{num14panero2}. The phase diagrams have common features of three transition lines separating the three phases, all meeting at a triple point.

The analytic treatment of the phase structure of noncommutative field theories has been based on the connection to matrix models, identified in \cite{steinacker05,PNT12}. Matrix models describing scalar field theory on the fuzzy sphere have been constructed both perturbatively \cite{ocon,samann,samann2} and non-perturbatively \cite{poly13} and models for other spaces are also known, such as fuzzy disc \cite{samanfuzzydisc} and fuzzy projective spaces \cite{JT14}. However, as we will see, these models are only approximations and a complete model is not yet known. Their phase structure has been studied and compared to the numerical results. Perturbative models are not well behaved close to the triple point of the theory, but the non-perturbative models give rise to phase structure similar to the one expected from the fuzzy field theory, even though so far the match has been a qualitative one, and not quantitative, see \cite{JT15APS} for a review.

The main goal if this paper is to work further on the analytical understanding of the phase structure of the scalar field theories on the fuzzy spaces. We develop a new technique to study such structure and extend the techniques used in \cite{JT15}, yielding a more accurate results for the phase diagram and the triple point for the fuzzy sphere case. The techniques are developed generally and are applicable to matrix models describing field theory on a general fuzzy space, and perhaps beyond.

As one of the results, which is sort of a by-product in this context, we get a very thorough phase structure of the asymmetric hermitian matrix model, in an extent not available in the current literature.

We first give the necessary material related to fuzzy field theories and hermitian matrix models in the section \ref{sec2}. We try to be brief, but also to present all the details that are used later in the text. We then analyze the asymmetric hermitian matrix model in the section \ref{sec3}. The phase diagram is given, with two phase transition lines meeting at a point. 
In the section \ref{sec4} we present a class of models that also give rise to an asymmetric configuration, but in a different mechanism. These models have multitrace action and are related to the fuzzy field theories. We present two different approaches to investigate their properties, one in a close connection to results obtained in the section \ref{sec3}.

Finally, we study the particular case of the fuzzy sphere in the section \ref{sec5} and present the phase diagram of the theory. We identify the location of the triple point, which is much closer to the most recent numerical studies than the previous analytical results.


\section{Fuzzy field theory, hermitian matrix models and connection of the two}\label{sec2}

In this section we review the connection between fuzzy field theories and matrix models. We introduce field theories on fuzzy spaces and we summarize the most important techniques used to analyze hermitian matrix models, both singletrace and multitrace. We then show how to construct a matrix model corresponding to a scalar field theory on a fuzzy space. 

\subsection*{Fuzzy field theory}

Fuzzy spaces are finite mode approximations to compact manifolds. The algebra of functions on such space is finite dimensional and the Fourier-like expansions of the fields into the momentum states consist of a finite number of terms \cite{steinacker_review}. Even though we obtain a certain discretization of the manifold in this way, there are no distinct points on the fuzzy space. It is rather divided into a finite number of cells with no sharp boundaries in between them, very much in an analogue of the phase space of the quantum mechanics. In this way the discretization preserves the full group of symmetries of the original manifold.

To define a field theory on the fuzzy space we first recall the case of a commutative field theory. The euclidean real scalar field theory on a commutative space is given by an action
\be 
S(\phi)=\int d^nx\,\slr{\half\phi\Delta\phi+\half m^2\phi^2+V(\phi)}\ ,
\ee
where $\Delta$ is the Laplace operator on the manifold and the integrals is over the whole $n$-dimensional space. Correlation functions are then defined by the functional path integral with the weight given by the action
\be\label{sec2_path}
\avg{\mathcal{O}(\phi)}=\frac{1}{Z}\int D\phi\,e^{-S(\phi)}\,\mathcal{O}(\phi)\ ,
\ee
where $Z$ is a normalization constant. In the fuzzy case we replace all quantities and operations by their noncommutative counterparts \cite{bal}. The exact realization of this depends on the particular space and will be given explicitly for the fuzzy sphere in the section \ref{sec5}, but in general a real field becomes a hermitian matrix, the laplacian becomes a double commutator operator acting on the matrix, the spacetime integral becomes a trace. So we have
\be\label{2fuzzyact}
S[M]=\mathcal{V}\ \trl{\half M \K M+\half r M^2+ V(M)}\ ,
\ee
where $\mathcal{V}$ is a volume factor which depends on the specific fuzzy space and $\K$ is the fuzzy Laplace operator. We have renamed the mass parameter to $r$, which is the standard notation in the matrix models setting.

Since the path integral over all field configurations in (\ref{sec2_path}) becomes an integral over the hermitian matrices, describing all correlation functions in the fuzzy field theory 
\be\label{sec2_matrix}
\avg{\mathcal{O}(M)}=\frac{1}{Z}\int dM\,e^{-S(M)}\,\mathcal{O}(M)
\ee
is equivalent to describing all expectation values in a random hermitian matrix model described by the probability distribution
\be\label{2fuzzyprob}
P(M)dM=e^{-S[M]}dM\ .
\ee

\subsection*{Essentials of hermitian matrix models}

We now describe some tools needed to analyze hermitian matrix models using the saddle point approximation. We will be quite explicit at certain points, since some of the expressions will be needed later in the text. For further details we reference the reader to some excellent reviews on this topic \cite{matrixmodels}. Let us start with singletrace matrix models.

{\bf Single trace models}

A single trace matrix model is an ensemble of ${N\times N}$ hermitian matrices with the probability distribution given by
\be\label{2_hermprob}
P(M)dM=e^{-N^2 S[M]}dM\ , dM=\prod_i dM_{ii}\prod_{i<j}d\textrm{Re}M_{ij}\,d\textrm{Im}M_{ij}\ ,
\ee
with the action given by\footnote{Note that this is the fuzzy field theory model (\ref{2fuzzyact}) without the kinetic term.}
\be\label{2act_single}
S[M]=\frac{1}{N}\trl{\half r M^2+ V(M)}\ .
\ee
We will get to the factors of $N$ shortly. Such action is invariant under the transformation ${M\to U\Lambda U^\dagger}$, where $\Lambda$ is the diagonal matrix of the eigenvalues of $M$ and ${U\in SU(N)}$. After such transformation (\ref{2_hermprob}) becomes
\be 
P(M)dM=e^{-N^2\lr{\half r c_2+\sum g_n c_n}}
\bigg(\prod_{i<j}(x_i-x_j)^2\bigg)
\bigg(\prod_i dx_i\bigg)dU\ ,
\ee
where ${V(x)=\sum_{n>2}g_n x^n}$, i.e. a polynomial potential, and where we have introduced the normalized moments
\be\label{sec2_moments_notation}
c_n=\frac{1}{N}\sum_{i=1}^N x_i^n\ .
\ee
The extra factor ${\prod (x_i-x_j)^2}$ is the Vandermonde determinant, which is a Jacobian of the transformation to the eigenvalues $\Lambda$ and the angular variables $U$. It can be exponentiated into the action and we obtain
\be 
P(M)dM=e^{-N^2\lr{\half r c_2+\sum_n g_n c_n-\frac{2}{N^2}\sum_{j<i}\log|x_i-x_j|}}\bigg(\prod_{i<j}(x_i-x_j)^2\bigg)\bigg(\prod_i dx_i\bigg)dU\ .\label{sec2_prob}
\ee
We will omit the $dU$ factor in the following, since for invariant quantities, it integrates to an uninteresting constant factor when computing averages (\ref{sec2_matrix}). We are left with an eigenvalue problem only.

The factors of $N$ have been introduced such that for large matrices the moments and the $\log$ term in the exponent are finite, or equivalently that the eigenvalues are of order $1$. To bring the action into this form might require a scaling of the parameters of the model.

As we increase $N$, the probability distribution for eigenvalues in (\ref{sec2_prob}) becomes concentrated on configurations which result in a small quantity in the brackets in the exponent. In the limit ${N\to\infty}$ the only eigenvalue configuration which contributes to the averages over eigenvalue configurations is the configuration which minimizes that expression. This is an Euler-Lagrange or saddle point condition, which using
\be\label{sec2_multider}
\pd{c_n}{x_i}=\frac{1}{N}\,n\, x_i^{n-1}
\ee
becomes
\be\label{2EL}
r x_i^E+V'(x_i^E)=\frac{2}{N}\sum_{j\neq i}\frac{1}{x_i^E-x_j^E}\ .
\ee
The large $N$ limit of averages is then an evaluation on the extremal configuration ${\{x_i^E\}}$ which solves this equation,
\be 
\avg{f(\{x_i\})}\overset{N\to\infty}{\longrightarrow}f(\{x_i^E\})\ .
\ee
To solve (\ref{2EL}) we first take the large $N$ limit which turns the eigenvalues into a continuous quantity $x$ and the sums over eigenvalues into integrals weighted by the eigenvalue distribution, i.e.
\be\label{2.13}
rx+V'(x)=2P\int dy\frac{\rho(y)}{x-y}\ , 
\ee
where ${P\int}$ denotes the principal value of an integral. Introducing a resolvent function
\be 
\omega(z)=\int dy\frac{\rho(y)}{z-y}
\ee
(\ref{2.13}) becomes a Riemann-Hilbert problem for a function of a complex variable $\omega(z)$ with a given discontinuity across a cut determined by the support of the eigenvalue distribution $\rho$ and analytic elsewhere \cite{complex}. Namely (\ref{2.13}) becomes
\be 
\omega(x+i0^+)+\omega(x-i0^+)=rx+V'(x)\ ,\ x\in\textrm{supp}\,\rho
\ee
and in terms of the resolvent the distribution $\rho$ is given by
\be \label{sec2_rho}
\rho(x)=-\frac{1}{2\pi i}\lr{\omega(x+i0^+)-\omega(x-i0^+)}\ .
\ee
Such function is given by
\be 
\omega(z)=\half\slr{rz+V'(z)-M(z)\sqrt{\sigma(z)}}\ ,
\ee
where ${M(z)}$ is a polynomial which does not change sign on connected components of the support of the distribution $\rho$ and is determined by the condition
\be\label{sec2_condOM}
\omega(z)\overset{|z|\to\infty}{\longrightarrow}\frac{1}{z}\ .
\ee
${\sigma(z)}$ is a polynomial which is negative on this support. To proceed further, we need to make an assumption about the support of the eigenvalue distribution, i.e. on the form of ${\sigma(z)}$. The two relevant cases will be a distribution supported on a single interval ${(a,b)}$ with ${\sigma=(z-a)(z-b)}$ and a distribution supported on two intervals ${(a,b)\cup (c,d)}$ with ${\sigma=(z-a)(z-b)(z-c)(z-d)}$. The endpoints of the interval will be determined by the condition (\ref{sec2_condOM}).

An important quantity in our discussion will be the free energy, i.e. the log of the partition function
\be
\F=-\frac{1}{N^2}\log\lr{\int dM\,e^{-N^2S[M]}}\overset{N\to\infty}{\longrightarrow}S(\rho)-\textrm{Vandermonde term}
\ee
which is
\be\label{sec2_freeGeneral}
\F=\int dx\,V(x)\rho(x)-\int dx\,dy\,\rho(x)\rho(x)\log|x-y|\ .
\ee
This quantity characterizes the probability of a given eigenvalue distribution.  If there is more than one solution to equation (\ref{2.13}) for given values of parameters then the solution with a lower free energy, i.e. the more probable solution, will be realized in the model.\footnote{We obtain only three conditions in the case of the two cut solution with support ${(a,b)\cup (c,d)}$. The fourth condition is supplied by this condition of the minimal free energy. We will get to this point later in section \ref{sec3}.}

{\bf Quartic model}

\begin{figure}%
\centering
\includegraphics[width=0.8\textwidth]{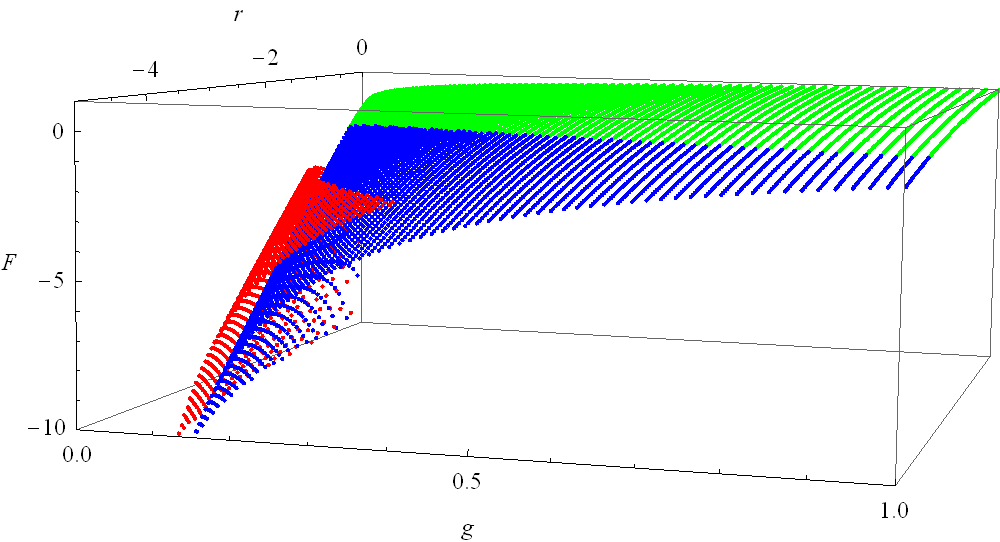}
\caption{Free energy diagram of the symmetric quartic hermitian matrix model (\ref{2act}). The green dots represent the symmetric one cut solution with the free energy (\ref{sec2_freeS1C}), the blue dots the symmetric two cut solution with the free energy (\ref{sec2_freeS2C}) and the red dots the asymmetric one cut solution with the free energy (\ref{sec2_freeA1C}). The transition between the green and the blue regions is given by the curve ${r=-4\sqrt{g}}$, the edge of the red region is given by the curve ${r=-2\sqrt{15}\sqrt{g}}$.}%
\label{fig_2freesym}%
\end{figure}

Without going into further details, let us present the solution for the symmetric quartic potential \cite{brezin,japonec}
\be\label{2act}
S[M]=\half r c_2+g c_4\ .
\ee
For ${r>-4\sqrt{g}}$ there is a symmetric one cut solution
\be\label{sec2_1Crho}
\rho(x)=\frac{1}{2\pi}\lr{r+2g \delta+4gx^2}\sqrt{\delta-x^2}\ ,
\ee
determined by
\be\label{2delta}
3\delta^2g+\delta r=4\ .
\ee
For ${r<-4\sqrt{g}}$ this solution becomes negative and the model suffers a phase transition\footnote{Which is of the third order.} into the two cut regime
\be
\rho(x)=\frac{2 g |x|}{\pi}\sqrt{\delta^2-(D-x^2)^2}\ ,
\ee
determined by
\be\label{2cond2cut}
4Dg+r=0\ ,\ \delta^2=\frac{1}{g}\ .
\ee
For ${r<-2\sqrt{15}\sqrt{g}}$ there exists a third solution, an asymmetric one cut solution
\be\label{2.26}
\rho(x)=\frac{1}{2\pi}\lr{2 \delta g + r  +4 D^2 g +  4 D g x+ 4 g x^2}\sqrt{\delta-\lr{x-D}^2}\ ,
\ee
determined by 
\begin{align}
2 D^2 g + 3 \delta g + \half r\,=\,&0\ ,\\
3 D^2 \delta g + \frac{3}{4} \delta^2 g + \frac{1}{4}\delta r\,=\,&1\ .\label{sec2_last}
\end{align}
The free energies of the three types of solutions are\footnote{The notation in this expression is hopefully clear. The subscript denotes the type of model, the superscript the type of solution in this model.}
\begin{align}
\F^{S\,1\textrm{ cut}}_{S}\,=\,&\frac{-r^2 \delta^2 + 40 r \delta}{384} - \half\log\lr{\frac{\delta}{4}} + \frac{3}{8}\ ,\label{sec2_freeS1C}\\
\F^{2\textrm{ cut}}_{S}\,=\,&-\frac{r^2}{16g} + \frac{1}{4}\log\lr{4g} + \frac{3}{8}\ ,\label{sec2_freeS2C}\\
\F^{AS\,1\textrm{ cut}}_{S}
=\,&-\frac{9}{8}-\frac{1}{4 \delta^2 g}-\frac{145}{64} \delta^2 g-\frac{75}{128} \delta^4 g^2-\half \log\lr{\frac{\delta}{4}}
\ .\label{sec2_freeA1C}
\end{align}
One can check that the two cut solution has always a lower free energy than the asymmetric one cut solution and thus is the preferred solution. The above free energies are plotted in the free energy diagram in the figure \ref{fig_2freesym}.

{\bf Multitrace models}

Opposing to the singletrace models such as (\ref{2act_single}), action of a multitrace model contains terms involving powers of a trace of some power of the matrix, e.g. $\lr{\trl{M^2}}^2$. Such terms introduce a further interaction of the eigenvalues. The large $N$ limit of these models is solved by converting them into an effective single trace model \cite{newcritical}. Let us give an example of a simple model with the action
\be
S[M]=c_2^2+\half r c_2+g c_4\ ,
\ee
to illustrate the idea, a general model is then a straightforward generalization. The saddle point equation for such model is
\be 2 c_2 x_i+r x_i+g x_i^3=\frac{2}{N}\sum_{i\neq j}\frac{1}{x_i-x_j}\ .
\ee
The first term can be viewed as an effective mass and this equation is the same, as an equation for a single trace model\footnote{This is analogue of the Hartree-Fock approximation and amounts to replacing ${\lr{\trl{M^2}}}^2$ by ${\avg{\trl{M^2}}\trl{M^2}}$.}
\be
S=\half \reff c_2+gc_4\ ,\ \reff=r+4c_2\ .
\ee
This model can be solved using (\ref{sec2_1Crho}-\ref{sec2_last}). We obtain an eigenvalue distribution with a known expression for the second moment. This moment however depends on $\reff$ which involves $c_2$, so the expression for $c_2$ becomes a selfconsistency condition determining $c_2$ in terms of the original parameter $r$. Solving this condition for $c_2$, we obtain also the effective parameter $\reff$ and thus the eigenvalue distribution with all the moments. The free energy of the solution is then given by the action and the Vandermonde term as before, but now includes an extra mutlitrace contribution compared to (\ref{sec2_freeGeneral}).

\subsection*{Matrix models of fuzzy field theories}

The fact that the fuzzy field theory is a particular hermitian matrix model stands as self-evident in the perspective of (\ref{sec2_matrix},\ref{2fuzzyprob}). Self-evident is however also the main obstacle in solving these types of models. The action (\ref{2fuzzyact}) is no longer invariant under ${M\to U\Lambda U^\dagger}$ diagonalization and the previously trivial group integral becomes an issue.

After the diagonalization we obtain for the probability density of the eigenvalues\footnote{We change the notation from $P(M)dM$ to $P(\Lambda)d\Lambda$ to stress that the angular integral has been done and the only degrees of freedom left are the eigenvalues.}
\be
P(\Lambda)d\Lambda=e^{-N^2\lr{\Seff(\Lambda)+\half r c_2+g c_4-\frac{2}{N^2}\sum_{j<i}\log|x_i-x_j|}}\prod_i dx_i\ ,
\ee
where we have introduced a kinetic term effective action defined \cite{steinacker05} by
\be\label{2angular}
\int dU\,e^{-\half \trl{U\Lambda U^\dagger[L_i,[L_i,U\Lambda U^\dagger]]}}=e^{-N^2 \Seff(\Lambda)}\ .
\ee
The effective action will depend on the eigenvalues of $M$ only, but in a very complicated way. At the moment, a complete formula for the integral on the left hand side of this equation is not known and only approximations are available.

One approach is to compute the integral perturbatively in the powers of $\Lambda$ \cite{ocon,samann,samann2}. The resulting expression for $\Seff$ is a multitrace model, which however does not behave well in the region of our interest in and we will not give more details about this technique here.

A different approach \cite{poly13} is based on the fact that the model corresponding to free theory ${g=0}$ is not too different from a pure matrix model \cite{steinacker05,PNT12} and the kinetic term in the action only rescales the distribution of the eigenvalues to $\sqrt\delta=2\sqrt{f(r)}$. Here 
\be\label{2f}
f(r)=\frac{1}{N}\sum_l \frac{n_l}{\K_l+r}\ ,
\ee
where $n_l$ is the multiplicity of the eigenvalue of $\K$ labelled by momentum $l$. Together with the translation invariance of the model this restricts the possible form of the effective action to be 
\be\label{2F}
\Seff=\half F[c_2-c_1^2]+\ldots\ .
\ee
The ellipsis stands for terms that give contributions to the saddle point equation which vanish when evaluated on the Wigner semicircle distribution. $F$ is a function to be determined as follows. Model (\ref{2F}) is a particular multitrace model and as such can be solved as described in the previous section. The radius of its semicircular distribution can be computed and matching it to the value ${2\sqrt{f}}$ fixes the function $F$ by
\be\label{2conditions}
F'[t]+z=\frac{N^2}{t}\ ,\ t=f(z)\ .
\ee
The second equation is to be inverted and the result to be plugged into the first equation. We will do this for the fuzzy sphere in section \ref{sec5}. In case $f$ is such that this can not be done, one has to resort to some different, perhaps perturbative solution of (\ref{2conditions}).

As an approximation to the integral (\ref{2angular}) one can then keep only the $F$ term in (\ref{2F}) and drop the rest of the terms. It turns out that the interesting features of the theory are in a region where this is a better approximation than the perturbative expansion of the angular integral \cite{JT14,JT15,JT15APS}. This is because such effective action is well behaved for both large and small values of $c_n$, as we will see in the section \ref{sec5}.

We will study this approximation in the rest of the text. As the first step, we will analyze a corresponding asymmetric single trace model.

\section{Asymmetric effective single trace model}\label{sec3}

In this section we focus on the most general quartic hermitian matrix model. Some aspects of such models have been considered before, see \cite{cm} for a review. Our quantitative results, most importantly the free energy diagram and the phase diagram in the figure \ref{fig_3phasediag}, are to our best knowledge new. 

Our main motivation to study this model is in the fuzzy field theory, but asymmetric quartic models have several other applications \cite{aplASQ1,aplASQ2,aplASQ3} and it would be interesting to see how to use these quantitative results there. We will use this model as one of the two tools to analyze fuzzy field theory matrix models, but the main results of that part of our paper can be understood without the results of this section.

\subsection*{Description of the model and different solutions}

The most general quartic action to be considered is (recall the notation (\ref{sec2_moments_notation}))
\be\label{3act}
S[M]=a c_1+\half r c_2+g c_4\ ,
\ee
since we can get rid of a cubic term by a shift in the matrix eigenvalues and a redefinition of the remaining parameters. We will later rescale the matrix to set the quartic coupling to unity, but for the moment we will keep the coupling constant explicit in our formulas.\footnote{Keep in mind though that a rescaling of the eigenvalues introduces a constant shift in the action due to the Vandermonde term in (\ref{sec2_prob}), which does not alter the saddle point equation for the distribution, but changes the free energy of the solution.}

We will also consider only $a>0$, since by taking ${x_i\to -x_i}$ we can set $a$ to be positive. The phase structure of the models with $a$ and $-a$ is the same, with the left and the right sides reversed.

Such model admits two different kinds of solution, one cut and two cut, both with an asymmetric support. Using the formulas (\ref{sec2_rho},\ref{sec2_condOM}) of the previous section, it is straightforward to write down the one cut solution supported on the interval ${\C=\lr{D-\sqrt{\delta},D+\sqrt{\delta}}}$ as
\be
\rho(x)=\frac{1}{2\pi}\lr{2 \delta g +r+4 D^2 g+4 D g x +4 g x^2}\sqrt{\delta-(D-x)^2}\ .
\ee
The two determining conditions for the endpoints of the interval are
\begin{align}
0\,=\,&\frac{1}{2}a+2 D^3g+3 D \delta g+\half D r\label{3det1}\ ,\\
1\,=\,&3 D^2 \delta g+\frac{3}{4} \delta^2 g+\frac{1}{4}\delta r\ ,\label{3det2}
\end{align}
and the moments of the distribution are given by
\begin{align}
c_1\,=\,&3 D^3 \delta g + \frac{3}{2} D \delta^2 g + \frac{1}{4}D \delta r\ ,\label{3c1}\\
c_2\,=\,&3 D^4 \delta g + 3 D^2 \delta^2 g + \frac{1}{4}\delta^3 g + \frac{1}{4} D^2 \delta r + \frac{1}{16}\delta^2 r\ .\label{3c2}
\end{align}
After some algebra, we can obtain also the free energy (\ref{sec2_freeGeneral})
\be
\F^{1\textrm{ cut}}_{AS}=\frac{3}{4}-\frac{2 D^2}{\delta}+3 D^4 g-\frac{9}{2} D^2 \delta g-\frac{1}{4}\delta^2 g-\frac{3}{2} D^2 \delta^3 g^2-\frac{3}{128} \delta^4 g^2-\half\log\left(\frac{\delta}{4}\right)\ ,\label{3freeA1C}
\ee
where we have used (\ref{3det1}),(\ref{3det2}) to eliminate the explicit $a$ and $r$ dependence.

In general there are going to be two different one cut solutions to the model for negative $r$. One with a negative $D$ occupying the left minimum of the potential and, if the right minimum is deep enough, one with a positive $D$ living in the $x>0$ region. We will refer to these as the left and the right one cut solution respectively.

In a similar fashion, we arrive at the double cut solution supported on 
\be
\C=\big(D_1-\sqrt{\delta_1},D_1+\sqrt{\delta_1}\big)\cup\big(D_2-\sqrt{\delta_2},D_2+\sqrt{\delta_2}\big)
\ee
given by
\be\label{sec3_rho2cut}
\rho(x)=\frac{2|D_1+D_2+g x|}{i\pi}\sqrt{\slr{\delta_1-(D_1-x)^2}\slr{\delta_2-(D_2-x)^2}}\ ,
\ee
with the endpoints determined by the following three equations
\begin{align}
\half r+\slr{2\lr{D_1^2+D_2^2+D_1 D_2}+\lr{\delta_1+\delta_2}}g\,=\,&0\ ,\label{sec3_2cut1}\\
\frac{1}{2}a+2\slr{ D_1 \delta_1 +D_2 \delta_2 - D_1 D_2 (D_1+D_2)}g\,=\,&0\ ,\label{sec3_2cut2}\\
\slr{D_1^2\delta_1+D_2^2\delta_2+\lr{D_1^2-D_2^2}\lr{\delta_1-\delta_2}-D_1 D_2\lr{\delta_1+\delta_2}+ \frac{1}{4}\lr{\delta_1-\delta_2}^2 }g\,=\,&1\label{sec3_2cut3}\ .
\end{align}
Expressions for the moments are presented in the appendix \ref{aA}. The expression for the free energy in terms of $D_{1,2},\delta_{1,2}$ is not reasonably obtainable and we will have to compute $\F$ numerically.

Notice that there are three equations and four unknowns to be determined. The system is underdetermined and there might be a large class of solutions of the two cut type. The preferred solution will be the one with the lowest free energy (\ref{sec2_freeGeneral}), as the most probable one. However it is not possible to write this condition down as an equation, as we do not have an analytic formula for the free energy of the two cut solution (\ref{sec3_rho2cut}) and we will have to find a way how to treat multiple two cut solutions.

\subsection*{Free energy diagram}

To analyze the asymmetric model, let us set $g=1$ and work with the action
\be\label{3actG1}
S[M]=a c_1+\half r c_2+ c_4\ ,
\ee
simply because we want to draw free energy diagrams and three parameters are one too many. In previous work \cite{JT15,JT15APS}, parameter $a$ has been set to unity, but this choice makes it easier to compare with the symmetric regime at ${a=0}$.

Any hope for analytic treatment of the model is long lost, since equations for the one cut solution (\ref{3det1},\ref{3det2}) can not be solved and equations for the two cut solution can not even be written down. We therefore analyze the model numerically. We will scan through the ${(a,r)}$ plane and look for solutions for given values of the parameters.

The case of the one cut solution is not very difficult, we simply look for the solution of equations (\ref{3det1},\ref{3det2}) and if a solution exists, we compute the free energy using (\ref{3freeA1C}). The case of the two cut solution is much more difficult.

We proceed as follows. We first choose a value of ${\delta_2}$ and plug it into (\ref{sec3_2cut1}-\ref{sec3_2cut3}). We start with ${\delta_2=0}$, which is the left one cut solution and then gradually increase this value. We look for the solution of the equations, and if we find a solution, we compute its free energy by numerically evaluating the corresponding integral (\ref{sec2_freeGeneral}) using (\ref{sec3_rho2cut}).

As we scan trough the parameter space, we come across three different situations. For some values of parameters there simply is no two cut solution. For some values there is a range of two cut solutions with one particular having the lowest free energy. Finally, for some values there is a range of two cut solutions but the one having the lowest free energy is the one with ${\delta_2=0}$, i.e. the one cut solution.

\begin{figure}%
\centering
\includegraphics[width=0.49\textwidth]{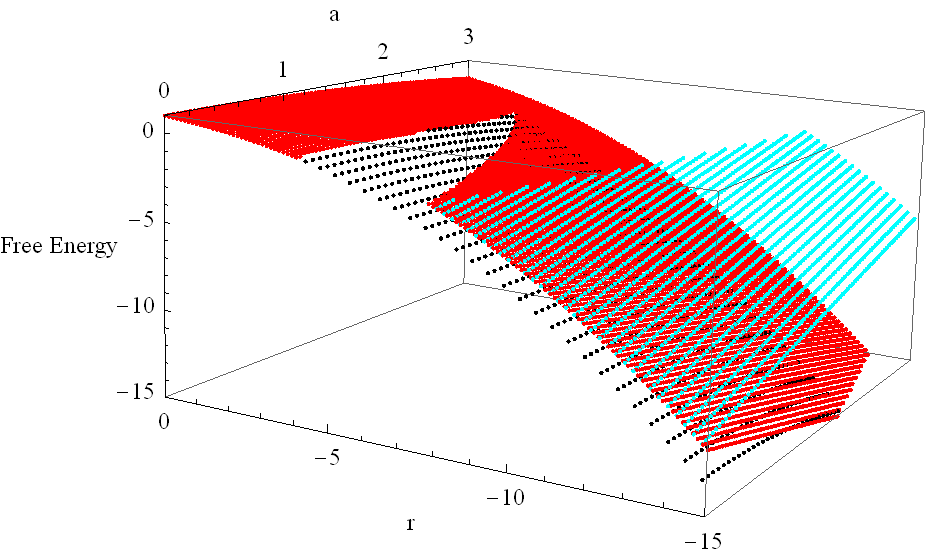}
\includegraphics[width=0.49\textwidth]{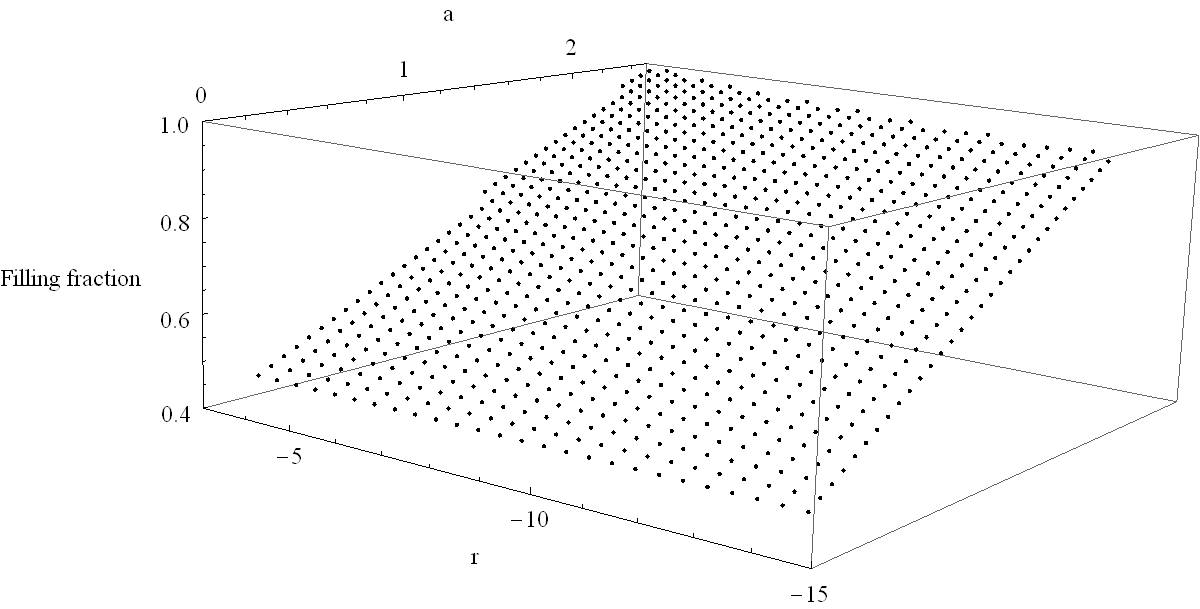}
\caption{Free energy diagram of the asymmetric quartic hermitian matrix model (\ref{3actG1}) and the filling fraction of the lowest free energy two cut solution. See the text for the description.}%
\label{fig_3freeAsym}%
\end{figure}

The resulting free energy diagram is given in the figure \ref{fig_3freeAsym}. The red dots denote the one cut solution with a negative $D$, the cyan dots denote the one cut solution with a positive $D$ and the black dots denote the proper two cut solution with the lowest free energy. For ${a=0}$, we obtain the expected structure of the symmetric model. One cut solution for ${r>-4}$ and two cut solution for ${r<-4}$. For values of $r$ less than ${-2\sqrt{15}}$ there exists another single cut solution, but with higher free energy as the double cut solution and the free energies of the left and the right one cut solution are the same. As we increase $a$, this structure remains roughly the same, but with the free energy gap between the two cut solution and the left one cut solution shrinking, with the right one cut solution gaining in energy. Finally, at certain point, the two cut solution with the lowest free energy is simply the left one cut solution and the situation remains like this with increasing $a$.

For a better intuition of what is happening in the model we give a plot of the free energy for a fixed value ${r=-12}$ and plots of eigenvalue distributions for different values of $a$ in the appendix \ref{aA}.

Figure \ref{fig_3freeAsym} also shows the filling fraction of the lowest energy two cut solution, i.e. the fraction of the eigenvalues located in the left minimum of the potential. It grows from $1/2$, i.e. a completely symmetric solution for $a=0$, to $1$ for some nonzero value of $a$, when the preferred solution becomes the one cut solution.

\subsection*{Phase transitions and phase diagram}

The merger of the two cut solution into the left one cut solution represents a phase transition of the model. From the free diagram we can see another transition line between the one cut and two cut solution and a third interesting line, which is a boundary of existence of the one cut solutions.

\begin{figure}%
\centering
\includegraphics[width=0.49\textwidth]{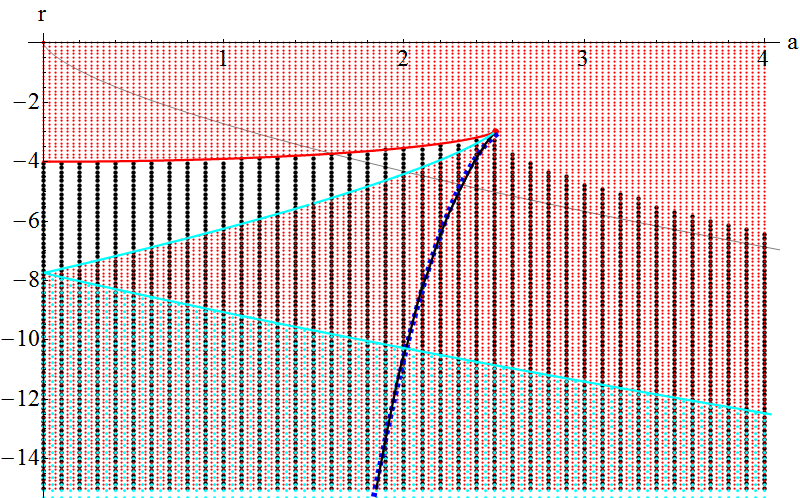}
\includegraphics[width=0.49\textwidth]{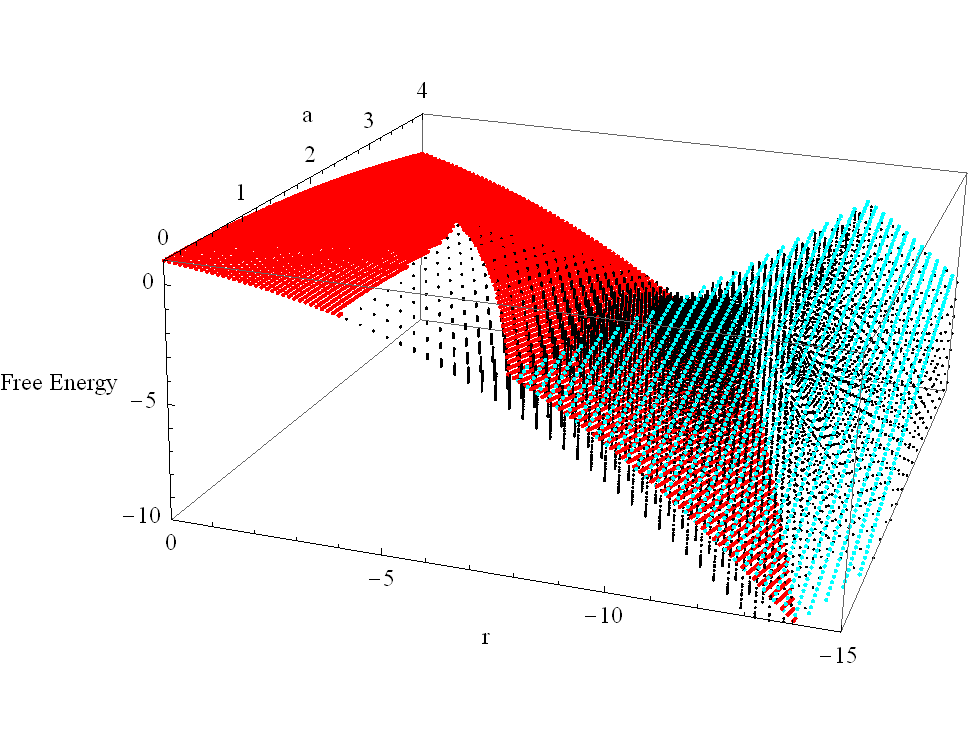}
\caption{Phase diagram of the asymmetric quartic hermitian matrix model (\ref{3actG1}) in the left image and the complete free energy diagram of the same model in the right figure. See the text for the description.}%
\label{fig_3phasediag}%
\end{figure}

The second and the third lines can be computed analytically, from the condition of existence of solution of the system (\ref{3det1},\ref{3det2}). The more important for us is the upper line separating the one cut and two cut regions, given by
\be\label{sec3_trafo1}
a^2=\frac{1}{135}\lr{22 r^3-480 r-(9r^3-80)^{3/2}}\ .
\ee
The other boundary of existence of the one cut solutions is given by roots of the polynomial (\ref{app_polynomial}) given in the appendix. It is a fourth order polynomial in $a^2$ so the solution can be found but the expression is way too complicated to present. Fortunately, we will need it just to check that our numerical procedure correctly reproduces the edge in the figure \ref{fig_3phasediag} and that this line and the line (\ref{sec3_trafo1}) terminate at the common point given by
\be\label{sec3_point}
(a,r)=\lr{\frac{32}{27} \sqrt{2\sqrt{5}},-\frac{4 \sqrt{5}}{3}}\ .
\ee
This point is the tip of the empty triangle region in the figure \ref{fig_3freeAsym}.

Finally, as the most important result of this section, we find the line where the two cut solution merges into the one cut solution. We do this using the same algorithm as outlined above, changing $\delta_2$ and looking for the solution with the lowest free energy. But we concentrate on a small region close to the expected transition line and run the calculation longer with a smaller step in $\delta_2$ to achieve a greater precision. We obtain a set of points giving the phase transition line. It turns out that for the region of our interest, the line is very well approximated by its best quadratic fit
\be\label{sec3_quadrfit}
r=-137.757+102.542\, a-19.4778\, a^2\ .
\ee
This formula however can not be a good approximation for $a$ all the way to ${a=0}$. In that case the two cut solution is preferred for any value of $r$. Since the free energy of the solutions is continuous with respect to $a$, there has to be a phase transition from the two cut solution to the one cut solution at some nonzero value of $a$ for any value of $r$.

The phase diagram of the model is given in the figure \ref{fig_3phasediag}. The small red dots denote points in the ${(a,r)}$ plane where a left one cut solution was found and the small cyan dots denote points, where a right one cut solution was found. The larger black dots denote points where a two cut solution exists. The solid red line denotes the phase transition line (\ref{sec3_trafo1}) and this is an analytical line. The cyan line denotes the edge of the existence region given by the roots of the polynomial (\ref{app_polynomial}). The solid black line denotes the phase transition line where a two cut solution with a lower free energy than the one cut solution ceases to exist, which is a numerical line. The largest red dot is the point (\ref{sec3_point}) where the two lines meet. The dashed blue line denotes the quadratic fit (\ref{sec3_quadrfit}) to the black line. Finally, the thin gray line separates the regions of parameter space, where the potential has a single minimum (above) and two distinct minima (below).

The numerical results are consistent with the line of transition from the two cut to the one cut regime (the black line) also terminating at the point (\ref{sec3_point}).

The phase diagram in the figure \ref{fig_3phasediag} consists of four different regions. The top red region, where only the left one cut solution exists. A triangular black region where only two cut solutions exist and a triangular red-black region where also a left one cut solution exists, but a particular two cut solution is the preferred one. Then there is a red-black region, where all the two cut solutions have higher free energy than the left one cut solution. Finally there are two red-black-cyan regions where both one cut solutions exist, together with all the two cut solutions in between them. In one of these regions a particular two cut solution is the preferred one, in the other it is the left one cut solution.

The figure \ref{fig_3phasediag} shows also the free energy diagram with all the possible two cut solutions and their free energies. They are not very important in this case, but when we will consider deformations of this free energy diagram in the next section, we will also have to consider these solutions.

We conclude this section by three notes. The two phase transitions from one cut to two cut solution are of the third order. Recall also that the diagram for ${a<0}$ is a mirror image of the diagram in the figure \ref{fig_3phasediag}. Finally, there is a region of parameter space, where a two cut solution exists even though there is only a single minimum in the potential and there is a region where this solution is preferred to the one cut solution. Conversely, there is also a region of parameter space (outside of the figure \ref{fig_3phasediag}), where a stable two cut does not exist even though there are two minima to the potential.

\section{Fuzzy-field-theory-like multitrace matrix models}\label{sec4}

We now present and investigate a class of multitrace models which will be referred to as fuzzy-field-theory-like multitrace models. Such models are given by the following multitrace action
\be\label{4model}
S[M]=\half F[c_2-c_1^2]+\half r c_2+g c_4\ ,
\ee
where ${F[t]}$ is some reasonably nice function of one variable. These matrix models naturally arise in the description of fuzzy scalar field theories, where the $F$-term is a contribution of the kinetic term, as described in the section \ref{sec2}. Let us recall that such action does not capture the field theory completely and is only an approximation.

\subsection*{The direct method}

We will now present a novel method to analyze the models (\ref{4model}). It will allow us to find solutions for given values of $r$ and $g$ and thus to control the region of the parameter space we study. As we will shortly see, this was not possible with the methods used previously in \cite{JT15}.

Using (\ref{sec2_multider}) we obtain a saddle point equation determining the eigenvalue distribution
\be\label{4saddle}
F'[c_2-c_1^2](x_i-c_1)+r x_i + 4 g x_i^3=\frac{2}{N}\sum_{j,j\neq i}\frac{1}{x_i-x_j}\ .
\ee
There are two different types of solutions to this model. One respects the ${M\to-M}$ symmetry of the original model (\ref{4model}) and we call solutions of this type the symmetric regime. Solutions of the other type break this symmetry and we call those the asymmetric regime. We analyze these regimes separately.

\subsubsection*{The symmetric regime}

In the symmetric regime all odd moments of the eigenvalue distribution vanish and the relevant effective model is (\ref{2act}). Looking at (\ref{4saddle}) we identify
\be\label{4_reff}
\reff=r+F'[c_2]\ .
\ee
The free energy of the symmetric solution is given by
\begin{align}
\F\,=\,&\half F[c_2]+\half r c_2 + g c_4-\int dx\,dy\, \rho(x)\rho(y)\log|x-y|=\no \,=\,&
\half F[c_2]+\half \lr{\reff-F'[c_2]} c_2 + g c_4-\int dx\,dy\, \rho(x)\rho(y)\log|x-y|=\no \,=\,&
\half\lr{F[c_2]-c_2 F'[c_2]}+\Feff\ ,\label{4_Free1}
\end{align}
where $\Feff$ is the free energy of the effective single trace model (\ref{sec2_freeS1C},\ref{sec2_freeS2C}). This formula is valid both for the one cut and the two cut solutions.

{\bf The symmetric one cut solution}

Together with (\ref{4_reff}) we have
\be\label{4_delta}
	\delta=\frac{\sqrt{\reff^2+48 g}-\reff}{6g}\ \ ,\ \ c_2=\frac{\delta}{4}+\frac{\delta^3 g}{16}\ .
\ee
The original spirit of these two equations is that for given values of parameters, we obtain $\delta,c_2$. We will turn this logic around, return back to the equation (\ref{2delta}) and consider it instead as an equation for $\reff$. After solving for $\reff$ in terms of $\delta$ we can get rid of $\reff$ in (\ref{4_Free1}) completely. Moreover we can also eliminate the $F'$ term using (\ref{4_reff}). We are left with just two equations which for given parameters $r$ and $g$ determine the eigenvalue distribution and free energy of the symmetric one cut solution. These are
\begin{align}
0\,=\,&\frac{4-3 \delta^2 g}{\delta}-r - F'\slr{\frac{4\delta + \delta^3 g}{16}}\label{4_cond1}\\
\F\,=\,&\frac{1}{4}+\frac{9}{128}\delta^4 g^2+\frac{1}{8}\delta r+\frac{1}{32} \delta^3 g r+\frac{1}{2} F\left[\frac{4\delta+\delta^3 g}{16}\right]-\frac{1}{2} \log\left(\frac{\delta}{4}\right)\label{4_freeS1C}
\end{align}

{\bf The symmetric two cut solution}

We can play the same game also with the two cut solution. For the two cut solution we have
\begin{align}\label{4_trafo}
	\reff=-4gD\ ,\, \delta=1/\sqrt{g}\ ,\ c_2=D\ ,
\end{align}
together with the formula (\ref{sec2_freeS2C}) for the free energy in the effective single trace model. The equations which determine the eigenvalue distribution and the free energy of the symmetric two cut solution to the model (\ref{4model}) are
\begin{align}
0\,=\,&4D g+r+F'\slr{D}\ ,\label{4_cond2}\\
\F\,=\,&\frac{3}{8}+D^2 g+\frac{D r}{2}+\frac{1}{2}F[D]+\frac{1}{4} \log\lr{4 g}\label{4_freeS2C}\ .
\end{align}

\subsubsection*{The asymmetric regime}

In the asymmetric regime the solution breaks the ${M\to-M}$ invariance of the original model and has nonzero odd moments. The relevant effective single trace model is thus
\be\label{4effaction}
\Seff[M]\,=\,\aeff  c_1+\half \reff c_2+ g c_4
\ee
with effective couplings given by
\begin{align}
\aeff \,=\,&-F'[c_2-c_1^2]c_1\ ,\\
\reff\,=\,&r+F'[c_2-c_1^2]\ .\label{4.5}
\end{align}
We will again use the same trick as in the symmetric regime and try to eliminate all the effective quantities from the formulas. We take (\ref{3det1},\ref{3det2}) and solve for $\aeff ,\reff$. Then we use this in (\ref{3c1},\ref{3c2}) to solve also for $c_1,c_2$. With these at our disposal we turn the equation (\ref{4.5}) into an equation for $\delta$
\be
4\frac{4+15\delta^2 g + 2 \delta r}{\delta(4+9\delta^2 g)}-F'\slr{\frac{\delta \lr{64  + 160 \delta^2 g+144 \delta^4 g^2+81 \delta^6 g^3+36 \delta^3 g r + 27 \delta^5 g^2 r}}{64(4+9\delta^2 g)}}=0\ .\label{4_cond3}
\ee

We obtain the expression for the free energy in the same fashion as in the symmetric case
\begin{align}
\F\,=\,&\half F[c_2-c_1^2]+\half r c_2 + g c_4-\int dx\,dy\, \rho(x)\rho(y)\log|x-y|=\no \,=\,&
\half F[c_2-c_1^2]+\half \lr{\reff-F'[c_2-c_1^2]} c_2 + g c_4-\int dx\,dy\, \rho(x)\rho(y)\log|x-y|=\no \,=\,&
\half\lr{F[c_2-c_1^2]-F'[c_2-c_1^2]c_2}-\aeff  c_1+\Feff=\no \,=\,&
\half\Big(F[c_2-c_1^2]-\lr{c_2-2c_1^2}F'[c_2-c_1^2]\Big)+\Feff\label{sec4effF2}\ .
\end{align}
Notice a crucial factor of $2$ next to $c_1^2$ in the $F'$ term. We now use the formula (\ref{3freeA1C}), eliminate the $F'$ term, use expressions (\ref{3c1},\ref{3c2}) for ${c_1,c_2}$ and obtain the final formula for the free energy of the asymmetric one cut solution
\begin{align}
\F\,=\,&-\frac{1}{128 g \left(4+9 \delta^2 g\right)^2}\Big[
6075 \delta^8 g^5
+3240 \delta^6 g^4 (4+\delta r)
+144 \delta^4 g^3 \left(29+40 \delta r+3 \delta^2 r^2\right)+\no
&+8 \delta^2 g^2 \left(-144+352 \delta r+117 \delta^2 r^2\right)+64 g \left(-8+8 \delta r+9 \delta^2 r^2\right)
+128 r^2\Big]+\no
&+\half F\slr{\frac{\delta \lr{64  + 160 \delta^2 g+144 \delta^4 g^2+81 \delta^6 g^3+36 \delta^3 g r + 27 \delta^5 g^2 r}}{64(4+9\delta^2 g)}}-\half \log\lr{\frac{\delta}{4}}\ .\label{4_free1AC}
\end{align}

Analytical solvability of the equations (\ref{4_cond1},\ref{4_cond2},\ref{4_cond3}) is clearly an issue, since for any nontrivial function $F$ we obtain a complicated set of equations. We will thus have to resort to a numerical solution.

This is however still a significant improvement over previous approaches, since we have to solve numerically only a single equation, which is done much faster. We also have control over the region of the parameter space we investigate, which was is not the case in the next method.

\subsection*{An alternative method}

There is an alternative method to study the free energy diagram of the matrix models (\ref{4model}), similar to the method used \cite{JT14,JT15}. Opposing to the direct computation presented above, this method uses the free energy diagram of the effective single trace model discussed in the section \ref{sec3} and deforms it.\footnote{The result obtained in \cite{JT15} used an effective model with ${a=1}$ and variable $g$. It also did not include all the various solutions possible for given values of the parameters.} These free energy diagrams are given by the figures \ref{fig_2freesym} and \ref{fig_3phasediag} for the symmetric and asymmetric regimes respectively. The idea is to take a point of the effective single trace free energy diagram ${(\reff,\geff,\Feff)}$ and find a corresponding point in the free energy diagram of the multitrace model. This is a point ${(r,g,\F)}$ for which the eigenvalue distribution is determined by the eigenvalue distribution given by $\reff$ and $\geff$.

For the symmetric regime the transformation is straightforward. We take a point ${(\reff,g,\Feff)}$ in the free energy diagram in the figure \ref{fig_2freesym}, and map it onto a point ${(r,g,\F)}$ given by (\ref{4_reff},\ref{4_Free1}) and the relevant expressions for the second moment. Namely
\begin{align}
r\,=\,&\reff-F'\slr{\frac{\delta}{4}+\frac{\delta^3 g}{16}}\ ,\label{sec4_meth2first}\\
	\F\,=\,&\half\lr{F\slr{\frac{\delta}{4}+\frac{\delta^3 g}{16}}-\lr{\frac{\delta}{4}+\frac{\delta^3 g}{16}}F'\slr{\frac{\delta}{4}+\frac{\delta^3 g}{16}}}+\frac{-\reff^2\delta^2+40 \reff \delta}{384}-\half \log\lr{\frac{\delta}{4}}+\frac{3}{8}\ ,\\ &\delta=\frac{\sqrt{\reff^2+48 g}-\reff}{6g}	\ ,
\end{align}
if $\reff>-4\sqrt g$ and
\begin{align}
r\,=\,&\reff-F'\slr{-\frac{\reff}{4g}}\ ,\\
	\F\,=\,&\half\lr{F\slr{-\frac{\reff}{4g}}+\frac{\reff}{4g}F'\slr{-\frac{\reff}{4g}}}-\frac{\reff^2}{16g}+\frac{1}{4}\log\lr{4g}+\frac{3}{8}\ ,\label{sec4_meth2middle}
\end{align}
if $\reff<-4\sqrt g$.

For the asymmetric regime the transformation is less straightforward. To bring the action (\ref{3actG1}) into the form (\ref{4effaction}) we need to rescale the eigenvalues of the matrix by a factor $g^{1/4}$. This means that the eigenvalue distribution in the multitrace model $\rho$ is related to the eigenvalue distribution in the effective model by
\be\label{sec4_altmethod1}
\rho(x)=g^{1/4}\rhoeff\Big(g^{1/4}x\Big)
\ee
and the moments are given by
\be\label{sec4_altmethod2}
c_n=\frac{1}{g^{n/4}}\cneff\ .
\ee
The rescaling introduces a factor of $g^{1/4}$ next to the linear term and $\sqrt{g}$ next to the quadratic term and identifying them with the coefficients in the model (\ref{4effaction}) we need to have
\begin{align}\label{sec4_altmethod3}
\begin{array}{l}
\aeff g^{1/4}\,=\,-c_1F'[c_2-c_1^2]\\
\reff\sqrt g\,=\,r+F'[c_2-c_1^2]
\end{array} \ \Rightarrow\ 
\begin{array}{l}
\aeff +\frac{\coneeff}{\sqrt g}F'\slr{\frac{\ctwoeff-\coneeff^2}{\sqrt g}}=0\\
r=\reff\sqrt g-\aeff \frac{\sqrt g}{\coneeff}
\end{array}\ .
\end{align}
The first equation is to be viewed as an equation for $g$ with known quantities of the effective model, the second equation then determines $r$. We can then compute the free energy pretty much in the same fashion as we computed in (\ref{sec4effF2}).
\begin{align}
\F\,=\,&\half \lr{F\slr{\frac{\ctwoeff-\coneeff^2}{\sqrt g}}-\frac{\ctwoeff-2\coneeff^2}{\sqrt g}F'\slr{\frac{\ctwoeff-\coneeff^2}{\sqrt g}}}+\Feff+\frac{1}{4}\log g=\\=\,&
\half\lr{F[c_2-c_1^2]-(c_2-2c_1^2)F'[c_2-c_1^2]}+\Feff+\frac{1}{4}\log g\ .\label{sec4_meth2last}
\end{align}
Note the extra log term compared to equations (\ref{4_Free1},\ref{sec4effF2}). This is a consequence of the scaling of the eigenvalues and the free energy of the effective asymmetric model having been computed at a unit coupling.

Finally, let us note that in this second method one has to consider all the possible solutions to the asymmetric effective model, since the transformation changes the free energy of the solution. It is possible that a solution with a higher free energy in the effective model has a lower free energy in the deformed diagram. These solutions were not available in \cite{JT15}, so this completes that analysis.

We have presented the methods to be used and we are now ready to proceed to the analysis of the model representing the scalar field theory on the fuzzy sphere.

\section{The case of the fuzzy sphere}\label{sec5}

In this section we use the method outlined in the section \ref{sec2} to obtain the matrix model corresponding to the scalar field theory on the fuzzy sphere and then analyze this model using techniques developed in the section \ref{sec4}. We will be brief in reviewing the material and mention only the concepts relevant for what we need. A curious reader is referenced to more thorough reviews \cite{steinacker_review,bal} or some of the original papers \cite{sF21,sF22}.

\subsection*{The fuzzy sphere}

The fuzzy sphere $S_F^2$ is a finite mode approximation to the regular sphere \cite{sF21}. It is constructed in terms of the algebra of functions, which is generated by the three coordinate functions ${x_i,i=1,2,3}$ obeying the following conditions
\be
\sum_{i=1}^3 x_i=R^2\ ,\ [x_i,x_j]=i\theta\ep_{ijk}x_k\ .
\ee
Here, $R$ is the radius of the sphere and $\theta$ is the noncommutativity parameter. These conditions can be fulfilled by representing $x$'s as generators of the $SU(2)$ in the ${N=2j+1}$ dimensional representation, which we denote ${L_i,i=1,2,3}$. Namely
\be
x_i=\frac{2 R}{\sqrt{N^2-1}}L_i\ ,\ \theta=\frac{2 R}{\sqrt{N^2-1}}\ .
\ee
This choice of coordinate functions keeps the radius of the sphere independent of $N$. The value of $R$ will not be important in what follows and we will set it to unity. We can see that in the large $N$ limit $\theta$ vanishes and we recover the regular sphere.

Real fields, which are elements of the algebra of functions, are arbitrary polynomials in $x$'s, i.e. general ${N\times N}$ hermitian matrices. Noncommutative analogue of the derivative is the $L$-commutator and the kinetic term of the field theory becomes
\be
\K M=\frac{1}{R^2}\sum_{i=1}^3[L_i,[L_i,M]]\ .
\ee
Eigenvalues of $\K$ are $j(j+1)$ with multiplicities $2j+1$ and thus the function $f$ in (\ref{2f}) becomes
\be
f(z)=\frac{1}{N}\sum_{j=0}^{N-1}\frac{2j+1}{j(j+1)+r}\ \to \int_0^1 dx\,\frac{2Nx}{N^2x^2+r}=\frac{1}{N}\log\lr{1+\frac{N^2}{r}}\ .
\ee
It is straightforward to plug this expression into (\ref{2conditions}), invert the function $f$ and compute $F$, obtaining the following result
\be\label{5sphereF}
F[t]=N^2\log\lr{\frac{t}{1-e^{-t}}}\ .
\ee
This is a growing function of $t$ and thus introduces an attractive interaction among the eigenvalues. It however grows only logarithmically and thus does not overcome the Vandermonde repulsion completely. Such attractive force can stabilize the asymmetric solution (\ref{2.26}) since it can be energetically beneficial for the eigenvalues to stick together in a higher potential rather than to split far apart. And as we will see, it will indeed be the case for some values of $r$ and $g$.

\subsection*{The free energy diagram}

We now have all the pieces in place for our main goal, the free energy diagram of the scalar field theory on the fuzzy sphere. We will obtain the diagram in two different ways outlined in the previous section.

\begin{figure}%
\centering
\includegraphics[width=0.49\textwidth]{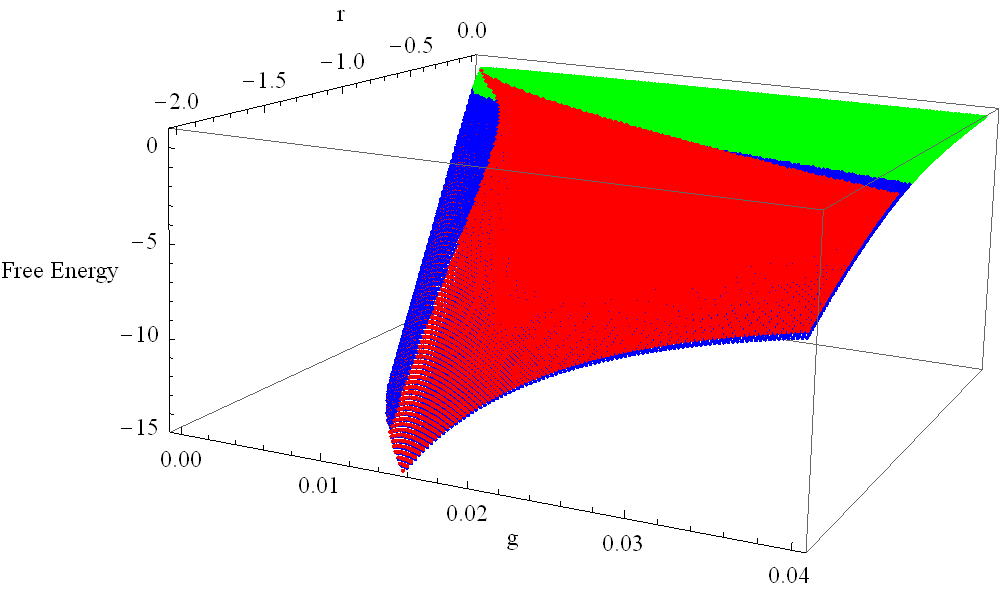}
\includegraphics[width=0.49\textwidth]{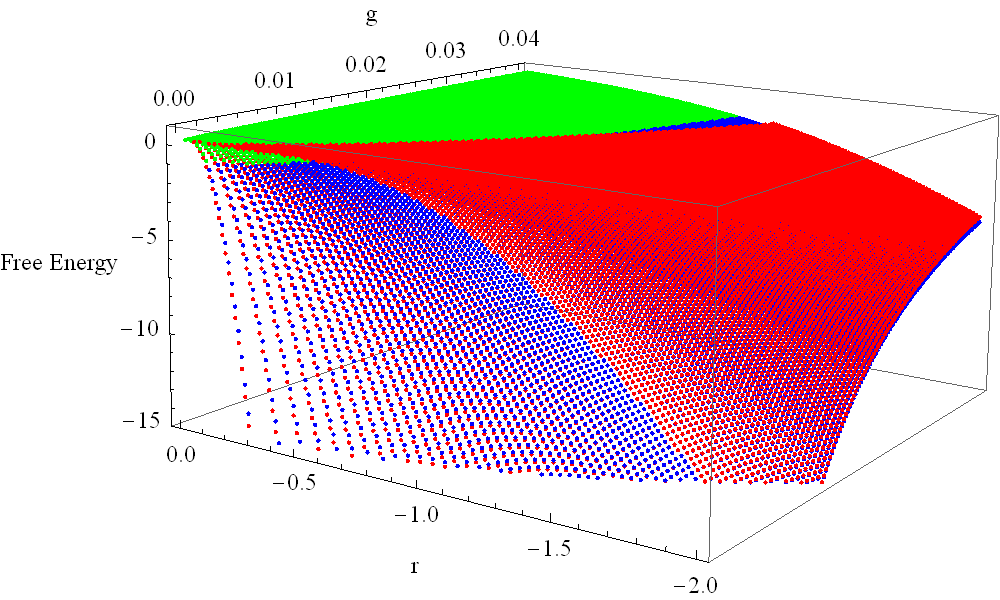}
\includegraphics[width=0.49\textwidth]{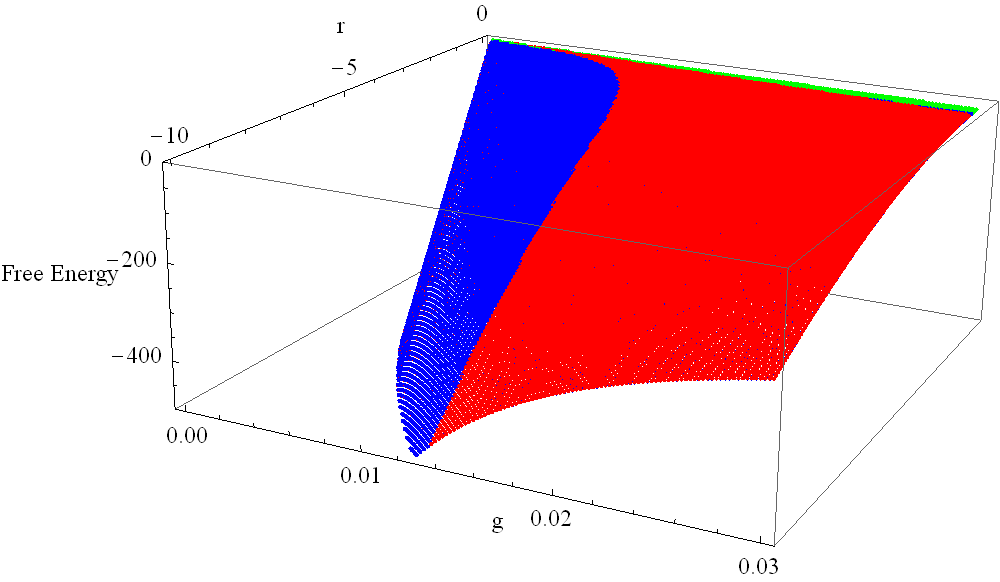}
\includegraphics[width=0.49\textwidth]{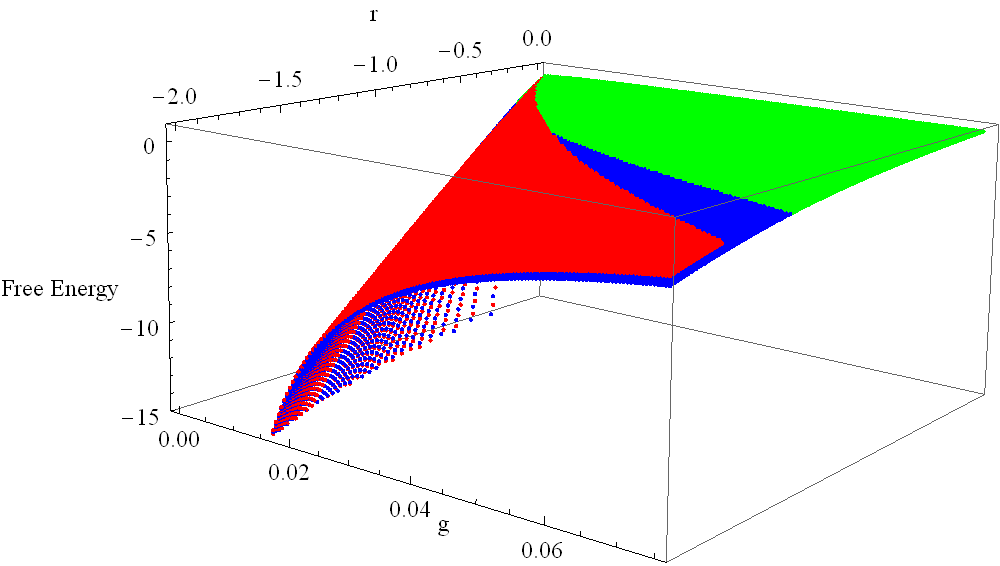}
\caption{Four different views of the free energy diagram of the matrix model corresponding to the scalar field theory on the fuzzy sphere, computed by the direct method. Note a different scale in the bottom two pictures. See the text for description.}%
\label{fig_5free1}%
\end{figure}

First, we use the direct computation. The expression (\ref{5sphereF}) for the function $F$ gives no hope for any analytical treatment and we will have to solve relevant equations numerically. We scan through the $(r,g)$ parameter space and look for solutions of equations (\ref{4_cond1},\ref{4_cond2},\ref{4_cond3}). If we find a solution, we compute the free energy, make a dot in the diagram and proceed further. As a result, we obtain the free energy diagram as in the figure \ref{fig_5free1}.

For a better understanding of what is happening in the model we present a slice of the free energy diagram for a fixed value of ${r=-0.5}$ in the appendix \ref{aA}. We also plot the eigenvalue distributions for several values of ${g}$.

We can see that the symmetric regime of the model is not very different from the symmetric regime of the simple quartic model (\ref{2act}) in the figure \ref{fig_2freesym}. However the behaviour of the asymmetric regime changes significantly. First of all the region of existence of the asymmetric one cut solution is shifted and extends also into the region of existence of the symmetric one cut solution. Moreover, there is a region where the free energy of the asymmetric solution is lower than in the symmetric case and in this region of parameter space the asymmetric solution is the preferred one.

The transition line between the two symmetric solutions, i.e. the line separating the green and the blue regions in the free energy diagram, can be computed analytically. In (\ref{4_trafo}) we take $\reff=-4\sqrt{g}$ which in (\ref{4_cond2}) yields
\be\label{5_line1}
r=-5\sqrt{g}-\frac{1}{1-e^{1/\sqrt{g}}}\ .
\ee
Moreover, a perturbative expression for the edge of the asymmetric one cut solution has been obtained in \cite{JT14} and the first few terms of the large $g$ expansion are given by
\be\label{5_line2}
r=-2 \sqrt{15} \sqrt{g}+\frac{2}{5}-\frac{19}{18000 \sqrt{15} \sqrt{g}}+\frac{11}{150000 g}+\frac{44373739}{1458000000000 \sqrt{15} g^{3/2}}+\ldots\ ,
\ee
with the rest of the formula given in the appendix. The lines in the figure \ref{fig_5free1} correctly reproduce both of these lines.

We now proceed to the second method of the section \ref{sec4} and obtain the diagram as a deformation of the effective single trace diagrams. For the symmetric regime it is straightforward to use formulas (\ref{sec4_meth2first}-\ref{sec4_meth2middle}) and we obtain a free energy diagram which perfectly matches the green and blue regions in the figure \ref{fig_5free1}. This serves as a very important independent check of the diagram.

For the asymmetric regime, we need to use the formulas (\ref{sec4_altmethod3}). To make sure that the results are as independent as possible from the previous method, we do not use the analytic formula (\ref{sec4_meth2last}) for the free energy but compute it by numerical integration directly from the eigenvalue distribution
\be
\F=\half F\slr{c_2-c_1^2}+\half r c_2+gc_4-\int dx\,dy\, \rho(x)\rho(y)\log|x-y|\ .
\ee
The eigenvalue density is determined by the expression (\ref{sec4_altmethod1}) and the moments are also computed numerically.

We now need to go through all the possible stable asymmetric solutions, since the multitrace term can in principle make previously unfavourable solution a preferred one. After a careful analysis we however find out that it is not the case. The transformation of the diagram in the figure \ref{fig_3phasediag} is quite dramatic, for example the right one cut solution is completely washed away and there are indeed cases of an asymmetric two cut solution becoming energetically preferred to the asymmetric one cut solution. But this all happens in a region where the symmetric regime dominates and at the end of the day no asymmetric two cut solution takes over. It would be interesting to analyze the deformation induced by the function (\ref{5sphereF}) in more detail, but as our main interest is elsewhere, we will proceed further in our original direction.

It has been much easier to study the full zoo of the asymmetric solutions to the model with the second method and even though these solutions do not contribute to the final free energy diagram, it was important to check that this is indeed the case. The second method however has a drawback, namely the lack of control over where in the new diagram the deformation carries the points of the effective diagram. It is very difficult to study a particular desired region of the ${(r,g)}$ space. Therefore the diagram obtained by this method is not well suited for the study of the asymmetric transition line and we refrain from showing the results of the second method in a figure. The most important message is that the two methods yield the same free energy diagrams.

Encouraged by this, we proceed to determine of the asymmetric phase transition line and the location of the triple point from the free energy diagram in the figure \ref{fig_5free1}.

\subsection*{The asymmetric phase transition and the triple point}

From the figure \ref{fig_5free1} we see that the symmetric solutions are not the preferred solution to the model everywhere. For larger values of $g$ the asymmetric solution has a larger free energy, similarly to the case of pure matrix model (\ref{2act}). However as we decrease the coupling the free energy of the asymmetric solution decreases faster and at a certain point it becomes lower than in the corresponding symmetric solution. The model enjoys a phase transition and these points form a crucial line in the phase diagram. It is beyond any hope to compute this line analytically and we will have to use numerical computation again.

\begin{figure}%
\centering
\includegraphics[width=0.49\textwidth]{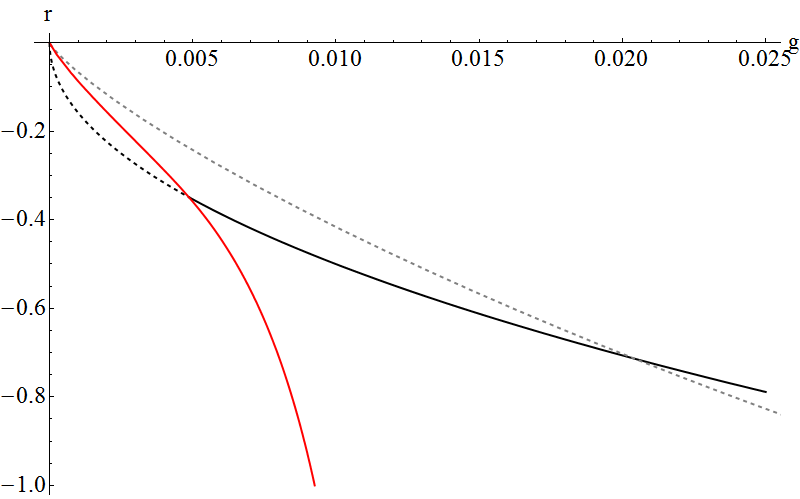}
\includegraphics[width=0.49\textwidth]{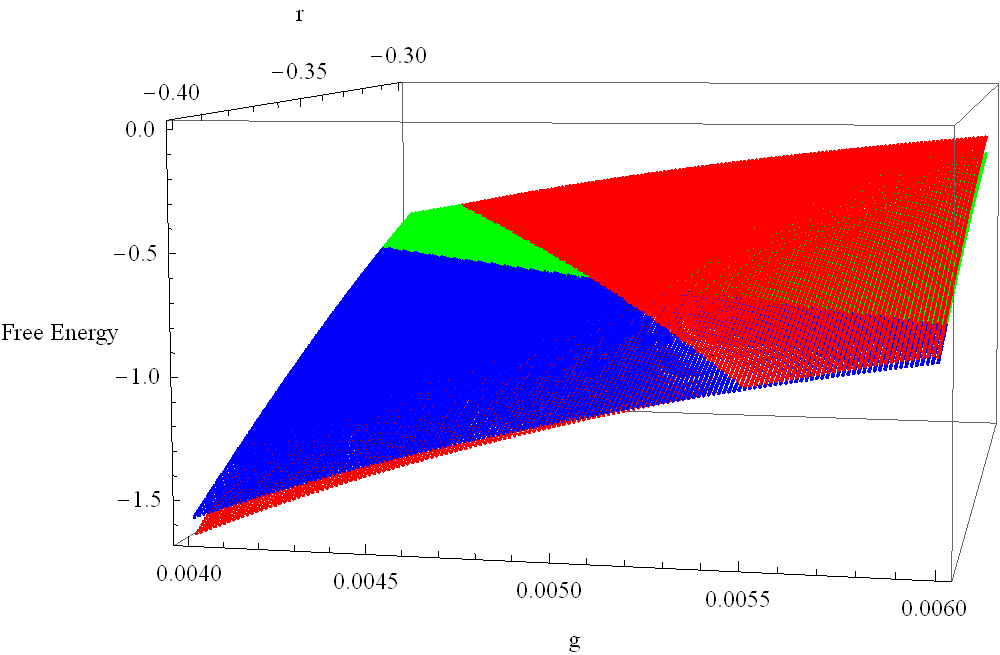}
\caption{The first image shows the phase diagram of the matrix model corresponding to the scalar field theory on the fuzzy sphere. The red line is the numerical line of the asymmetric phase transition. The black line is the analytical line of the symmetric phase transition (\ref{5_line1}), with the dashed part corresponding to the part which is not realized in the model. The gray dashed line represents the boundary of existence of the asymmetric solution. Intersection of this line with the symmetric phase transition line was previously identified as the location of the triple point.
The second image shows the free energy diagram very close to the triple point. The red dots denote the asymmetric one cut solution, the green dots the symmetric one cut solution and the blue dots the symmetric two cut solution.}%
\label{fig_5triplepoint}%
\end{figure}

We can compute the asymmetric phase transition line in two ways. We can either compute the intersection line of the three regions in the figure \ref{fig_5free1} by interpolating the surface between the numerical points. Or we can numerically solve the set of equations (\ref{4_cond1}),(\ref{4_cond3}) and (\ref{4_freeS1C})=(\ref{4_free1AC}) for the transition between the symmetric and the asymmetric one cut solutions and the set of equations (\ref{4_cond2}),(\ref{4_cond3}) and (\ref{4_freeS2C})=(\ref{4_free1AC}) for the transition between the asymmetric one cut solution and the symmetric two cut solutions. In both cases, we obtain a series of points, which represent the transition line numerically. And without much surprise the two approaches yield the same transition line, even though the first approach is much easier to employ due to complicated nature of the equations involved in the second approach. The final results are shown in the figure \ref{fig_5triplepoint}. 

We can see that the two transition lines intersect at a point where all three phases of the theory coexists. This is the triple point of the theory and the figure \ref{fig_5triplepoint} shows also the free energy diagram in the vicinity of this point. We can see that the transition extending from the triple point towards the origin of the diagram is the asymmetric transition line. Computing the location of the triple point is again a numerical task and we find the following value for the triple point of the model (\ref{4model},\ref{5sphereF})
\be\label{sec5_triplepoint}
\lr{g_c,r_c}\approx \lr{0.0049,-0.35}\ .
\ee
The numerical value of the critical coupling might seem a little small, but that is just a matter of the convention in (\ref{4model}). Any finite value is a significant difference from the original model (\ref{2act}) which does not have a stable asymmetric solution at all and has no triple point.

Let us now compare the free energy diagram in the figure \ref{fig_5free1}, the phase diagram in the figure \ref{fig_5triplepoint} and the location of the triple point (\ref{sec5_triplepoint}) to the available numerical results. Two previous numerical works \cite{num09,num14} have localized the triple point further away from the origin at around ${g_c \approx 0.15}$. Both these works however used linear extrapolation from regions quite far away and this could not capture some of the features of the transition lines. Also the size of the matrices has not been large enough to eliminate the finite $N$ effects. More thorough investigations showed, that the triple point is much closer to the origin \cite{denjoe_unpublished}. Our value of $g_c$ lies within the uncertainty interval of the most recent numerical simulations \cite{samo}. This is a surprisingly good agreement considering the approximation we have made. The value of $r_c$ differs roughly by a factor of $10$, which suggests that close to the triple point the neglected terms change the behaviour of the model in the $r$ direction but not in the $g$ direction. 

Before we conclude our report, let us note one last interesting property of the phase diagram of the model (\ref{4model},\ref{5sphereF}) in the figure \ref{fig_5triplepoint}, which is however easier seen from the third free energy diagram in the figure \ref{fig_5free1}. The asymmetric phase transition line takes a rather sharp turn with growing ${-r}$ towards the negative $r$ axis and for larger values runs almost parallel to it. This means that for very negative $r$ the phase transition happens for a coupling that changes very insignificantly. This is a very different from the behaviour of the existence line of the asymmetric solution, which takes a turn towards the $g$ axis and behaves as a square root of the coupling for large $r$. It is also a very different behaviour from the asymmetric phase transition lines obtained in the Monte Carlo simulations, which seem to be quite linear.

\section{Conclusions}\label{sec_conclusions}

To conclude, let is summarize our results. We have analyzed two different hermitian random matrix models. First, we have given a thorough analysis of an asymmetric quartic hermitian matrix model far beyond what is currently available. We have given the free energy diagram of the model and identified two phase transition lines. Both are one cut to two cut phase transitions. The first is a continuation of the original symmetric model phase transition in which the wells of the potential become too deep and the eigenvalue distribution splits. The second is unique to the asymmetric model and occurs when the difference between the two wells becomes too large and it is energetically more efficient for the eigenvalues to occupy the same minimum. The first line was obtained analytically, the second line was only possible to obtain numerically. The first line terminates at a point in the parameter space and the second line seems to begin there.

The second random matrix model we have analyzed is a model with a multitrace interaction mimicking the scalar field theory on fuzzy spaces. We have given two different methods to investigate this model, one relying on the asymmetric quartic matrix model and one doing independent calculations directly in the model. The first is an improved version of method used previously, the second is completely new. We have used these two methods to analyze the model corresponding to the theory on the fuzzy sphere.

We have identified the phase transition lines between the three phases of this model and we have localized the triple point. The structure of the phase diagram agrees with the expectations from the previous numerical investigations. The location of the triple point (\ref{sec5_triplepoint}) agrees qualitatively with previous numerical results and the critical coupling agrees well with a new work soon to be presented \cite{samo}. Our results also show a new interesting behaviour of the asymmetric phase transition far from the triple point.

One of the important questions for the further research is the impact of the neglected terms in (\ref{2F}). Their form is quite restricted and some of the terms are known exactly for the fuzzy sphere, but it is not clear how to treat them in a nonperturbative way. It would be interesting to see their role in the behaviour of the asymmetric transition line far from the origin of the parameter space and also in the suggested significance in the $r$ direction of the diagram, but not in the $g$ direction, at least close to the triple point.

A second aspect of the fuzzy field theory models that needs to be improved on is the understanding of the angular integral (\ref{2angular}). Complexity of the calculations needed for the fourth moment calculations mentioned above suggest, that in order to make reasonable progress in the understanding of the model (\ref{2fuzzyprob}) one needs to find a different description of the kinetic term effective action, rather than the multitrace calculations.

Numerical results for spaces beyond $S_F^2$ are available, such as the fuzzy disc \cite{num_disc} and ${\mathbb R\times S_F^2}$ \cite{num_RSF2}. It should be possible to use the presented approach to obtain the corresponding matrix models for these spaces, analyze their phase diagram and compare with the numerical results. Moreover, other spaces such as ${\mathbb CP^n_F}$ or the fuzzy four sphere $S_F^4$ are also promising candidates and would provide a prediction for future numerical simulations.

Recently, a new method has been proposed to carry out loop calculations in noncommutative field theories \cite{steinacker16} and it would be very interesting to try to approach this problem from this perspective looking carefully into the phase structure of noncommutative field theories directly in the terms of the field theory. This is especially interesting in the case of noncompact spaces, where the matrix model formulation is more involved.

Finally, clearly an analytical treatment of the asymmetric quartic hermitian matrix model would be very helpful to analyze the model further. One possibility is to find a perturbative solution to equations (\ref{3det1},\ref{3det2}) in powers of $a$ and then compute the free energies of the two solutions as a correction to the symmetric free energy. Perhaps one can go far enough in the perturbative series to obtain relevant information.

{\bf Acknowledgements.}\ 
I would like to thank Harold Steinacker, Denjoe O'Connor, Samuel Kov\'a\v cik, Eduard Batmendijn and M\'aria \v Subjakov\'a for many useful discussions.

I would also like to thank STP DIAS for hospitality and support during my visit in June 2016. This visit was also supported by the Action MP1405 QSPACE, supported by COST (European Cooperation in Science and Technology) within the 3rd STSM call. I have received travel support from the same action on numerous other occasion, which is also gratefully acknowledged.

This work was supported by the \emph{Alumni FMFI} foundation as a part of the \emph{N\'{a}vrat teoretikov} project and by VEGA 1/0985/16 grant.


\appendix
\section{Explicit formulae and additional plots}\labell{aA}
In this appendix we give several explicit formulae which were too lengthy to present in the main text.

The following are the first and the second moment of the asymmetric two cut solution (\ref{sec3_rho2cut}) to the model (\ref{3act})
\begin{align}
c_1\,=\,&2 D_1^3 \delta_1 g-D_1^2 D_2 \delta_1 g-D_1 D_2^2 \delta_1 g+D_1 \delta_1^2 g-D_1^2 D_2 \delta_2 g-\no
&-D_1 D_2^2 \delta_2 g+2 D_2^3 \delta_2 g-D_1 \delta_1 \delta_2 g-D_2 \delta_1 \delta_2 g+D_2 \delta_2^2 g\ ,\\
c_2\,=\,&2 D_1^4 \delta_1 g-D_1^3 D_2 \delta_1 g-D_1^2 D_2^2 \delta_1 g+\frac{9}{4} D_1^2 \delta_1^2 g-\frac{1}{4} D_1 D_2 \delta_1^2 g-\no&
-\frac{1}{4} D_2^2 \delta_1^2 g+\frac{\delta_1^3 g}{8}-D_1^2 D_2^2 \delta_2 g-D_1 D_2^3 \delta_2 g+2 D_2^4 \delta_2 g-D_1^2 \delta_1 \delta_2 g-\no&
-\frac{3}{2} D_1 D_2 \delta_1 \delta_2 g-D_2^2 \delta_1 \delta_2 g-\frac{1}{8} \delta_1^2 \delta_2 g-\frac{1}{4} D_1^2 \delta_2^2 g-\frac{1}{4} D_1 D_2 \delta_2^2 g+\no&
+\frac{9}{4} D_2^2 \delta_2^2 g-\frac{1}{8} \delta_1 \delta_2^2 g+\frac{\delta_2^3 g}{8}\ .
\end{align}

The following is the polynomial, which determines the lower edge of the region of existence of the left one cut solution in the figure \ref{fig_3phasediag} and the edge of the region of existence of the right one cut solution, i.e. the cyan line in the same figure. The line is given by the roots of the following expression
\begin{align}\label{app_polynomial}
19110297600-201553920 a^4+2657205 a^8+2149908480 a^2 r+955514880 r^2\no
+75582720 a^4 r^2+50761728 a^2 r^3+314928 a^6 r^3+1990656 r^4+52488 a^4 r^4\no
-186624 a^2 r^5-525312 r^6+13122 a^4 r^6+3888 a^2 r^7-5616 r^8+216 a^2 r^9
+72 r^{10}+r^{12}\ .
\end{align}

The following is the full perturbative expression for the boundary of existence of the asymmetric one cut solution (\ref{5_line2})
\begin{align}
r\,=\,&-2 \sqrt{15} \sqrt{g}
+\frac{2}{5}
-\frac{19}{18000 \sqrt{15} \sqrt{g}}
+\frac{11}{150000 g}
+\frac{44373739}{1458000000000 \sqrt{15} g^{3/2}}+\no&
+\frac{5033447}{6075000000000 g^2}
+\frac{90528767950213}{248005800000000000000 \sqrt{15} g^{5/2}}+\no&
+\frac{5907303225637}{516678750000000000000 g^3}
+\frac{25212604606236508759}{4464104400000000000000000000 \sqrt{15} g^{7/2}}
\ .
\end{align}

Finally we present several extra figures to illustrate the results of the report better. All the figures are described in the captions.

\begin{figure}[H]%
\centering
\includegraphics[width=0.9\textwidth]{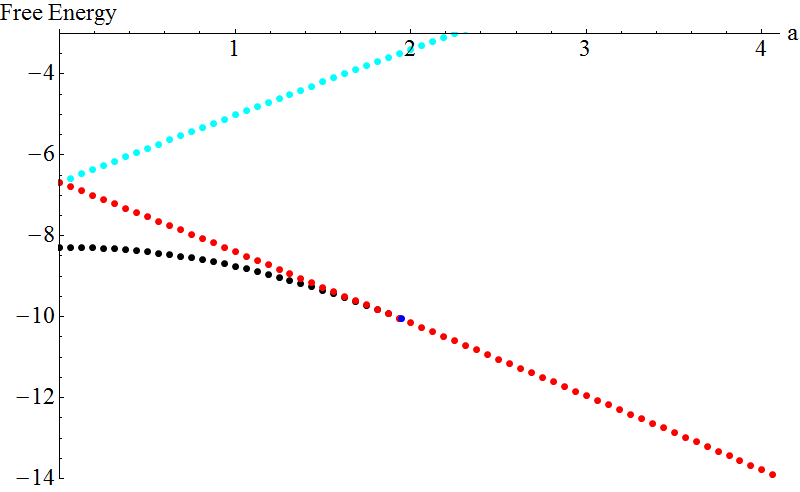}
\caption{A slice of the free energy diagram of the asymmetric quartic model (\ref{3actG1}) shown in the figure \ref{fig_3freeAsym} for ${r=-12}$. The red dots represent the left one cut solution, the cyan dots denote the right one cut solution, the black dots represent the two cut solution and the blue dot denotes the point of phase transition according to (\ref{sec3_quadrfit}).}%
\label{fig_B1}%
\end{figure}

\begin{figure}[H]%
\centering
\includegraphics[width=0.9\textwidth]{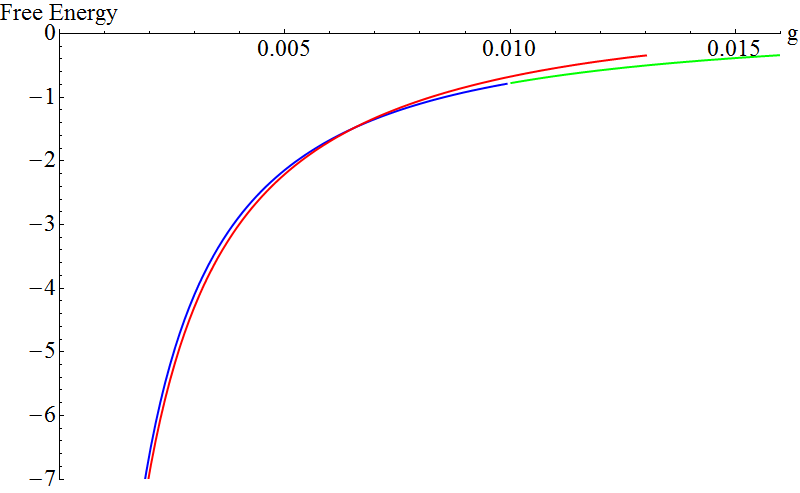}
\caption{A slice of the free energy diagram of the fuzzy sphere matrix model (\ref{4model},\ref{5sphereF}) shown in the figure \ref{fig_5free1} for ${r=-0.5}$. The dots have been connected for a better orientation. The red line is the energy of the asymmetric one cut solution, the blue line is the energy of the symmetric two cut solution and the green line is the energy of the symmetric one cut solution.}%
\label{fig_B2}%
\end{figure}

\begin{figure}[H]%
\centering
\includegraphics[width=0.32\textwidth]{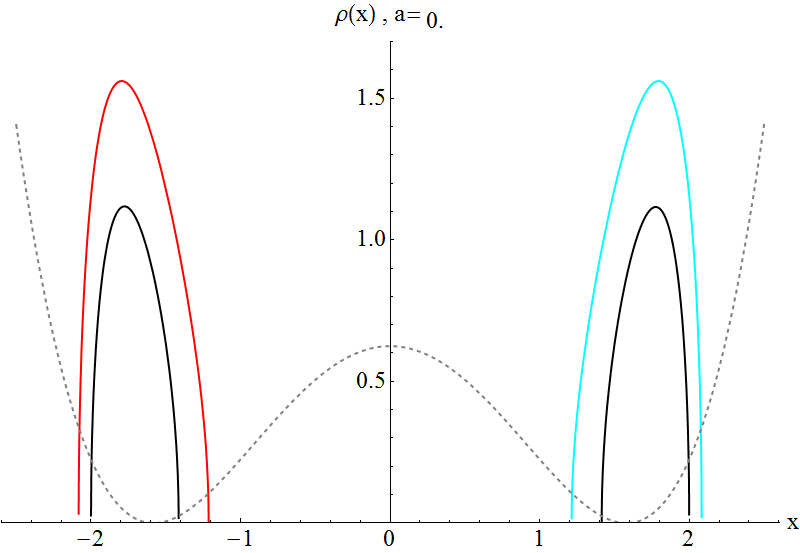}
\includegraphics[width=0.32\textwidth]{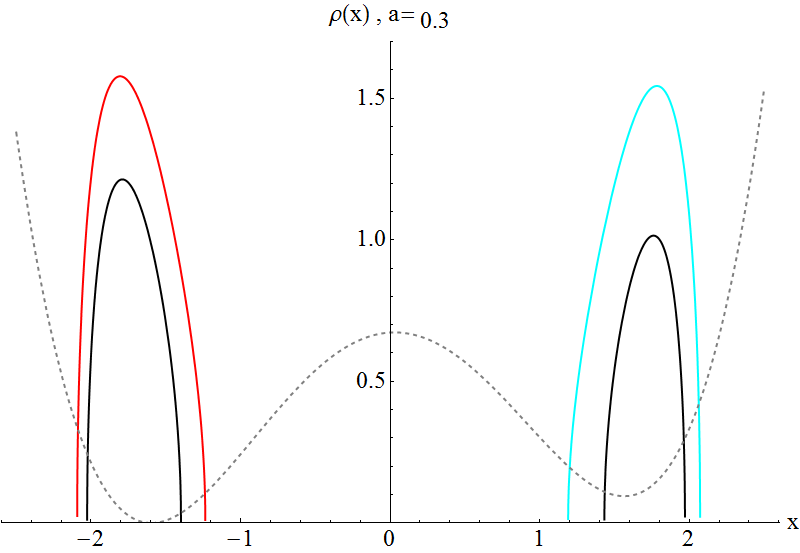}
\includegraphics[width=0.32\textwidth]{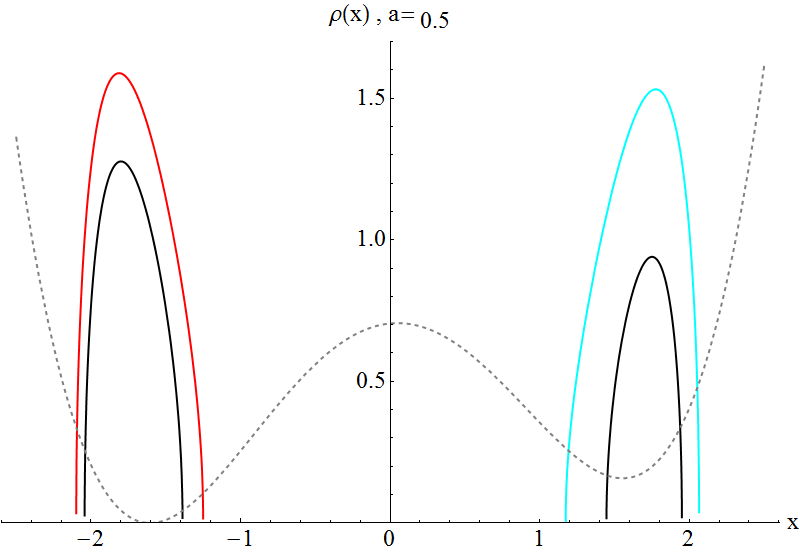}
\includegraphics[width=0.32\textwidth]{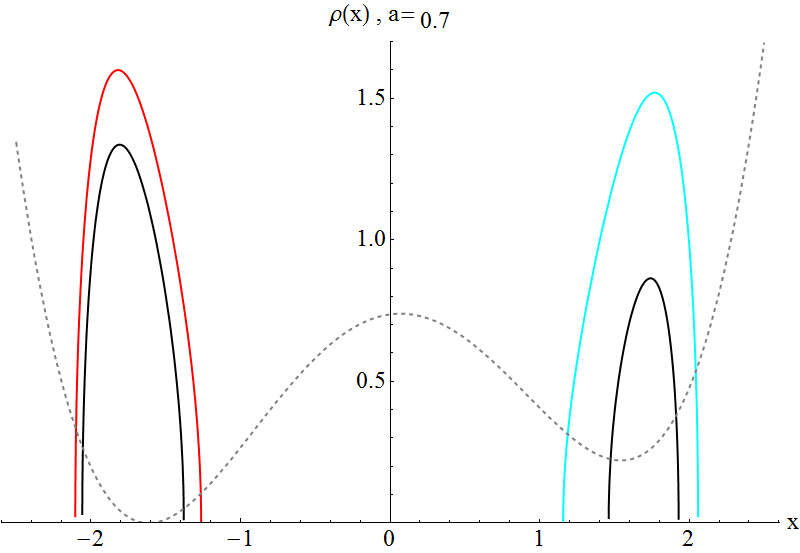}
\includegraphics[width=0.32\textwidth]{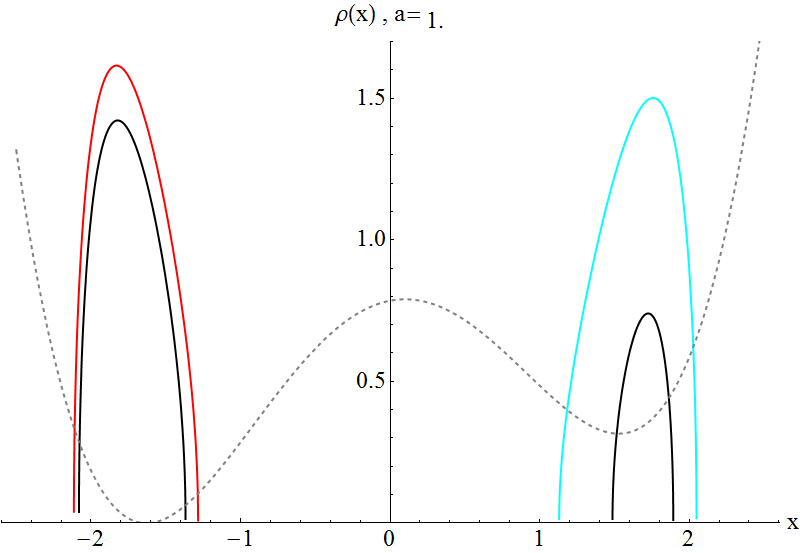}
\includegraphics[width=0.32\textwidth]{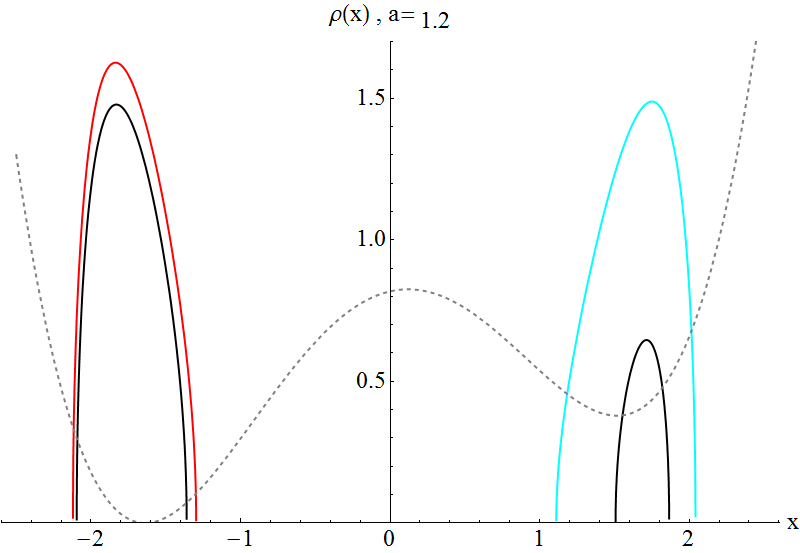}
\includegraphics[width=0.32\textwidth]{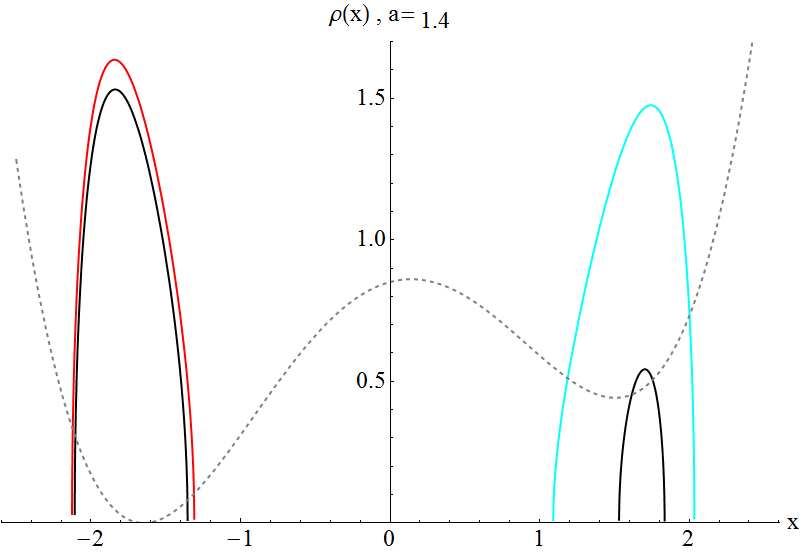}
\includegraphics[width=0.32\textwidth]{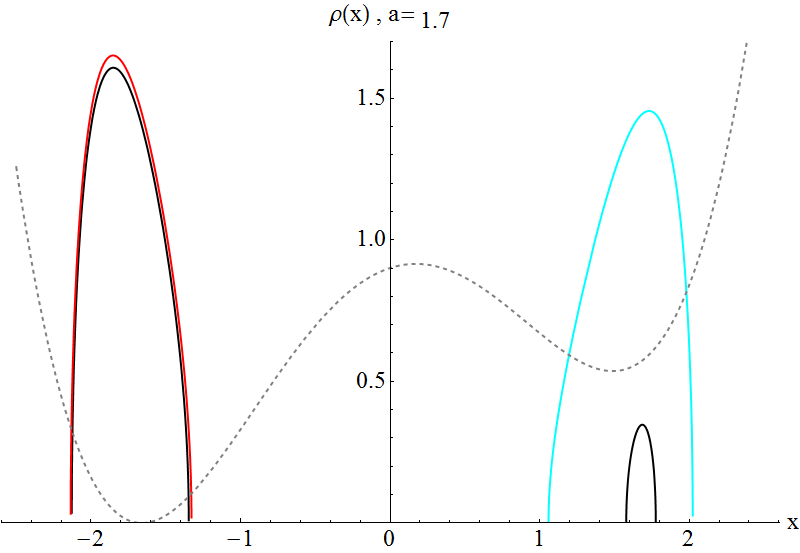}
\includegraphics[width=0.32\textwidth]{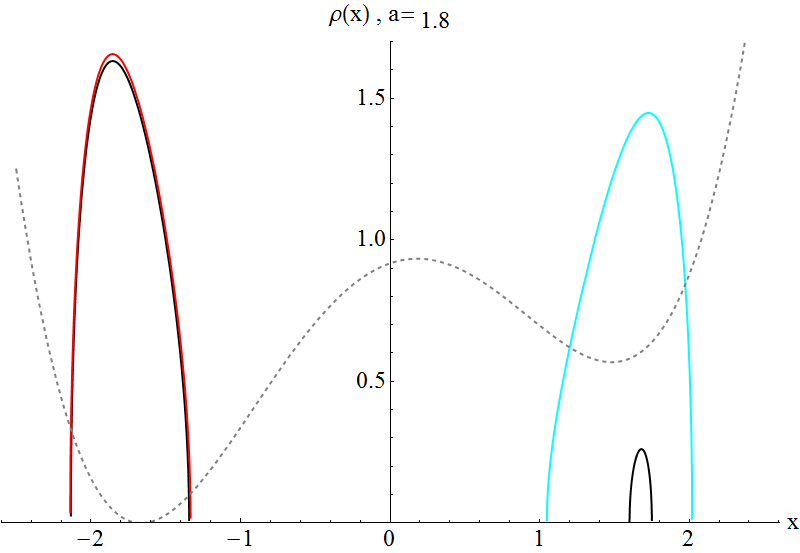}
\includegraphics[width=0.32\textwidth]{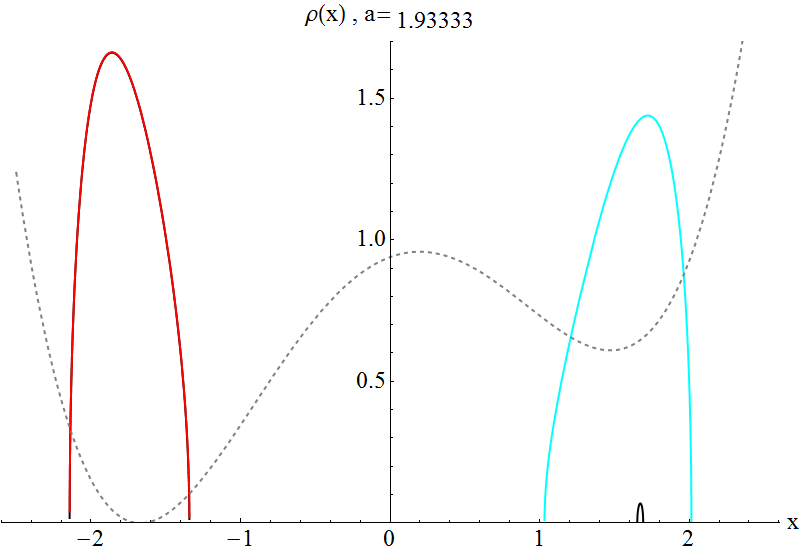}
\includegraphics[width=0.32\textwidth]{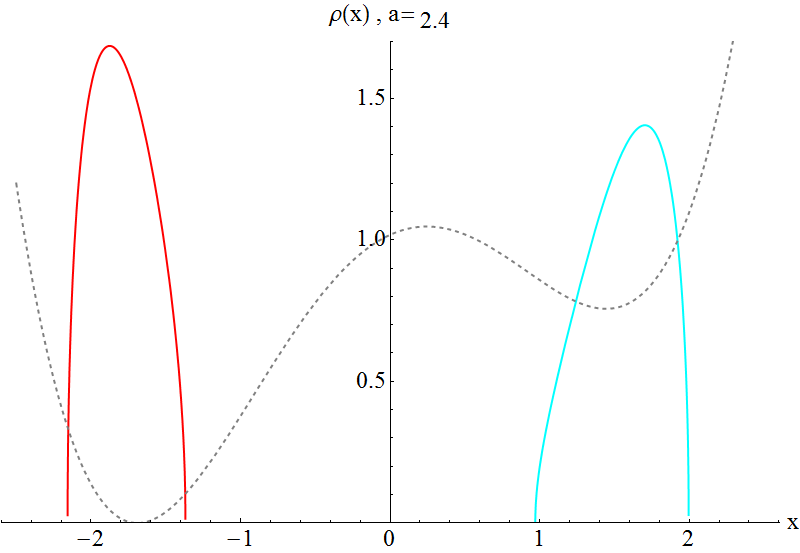}
\includegraphics[width=0.32\textwidth]{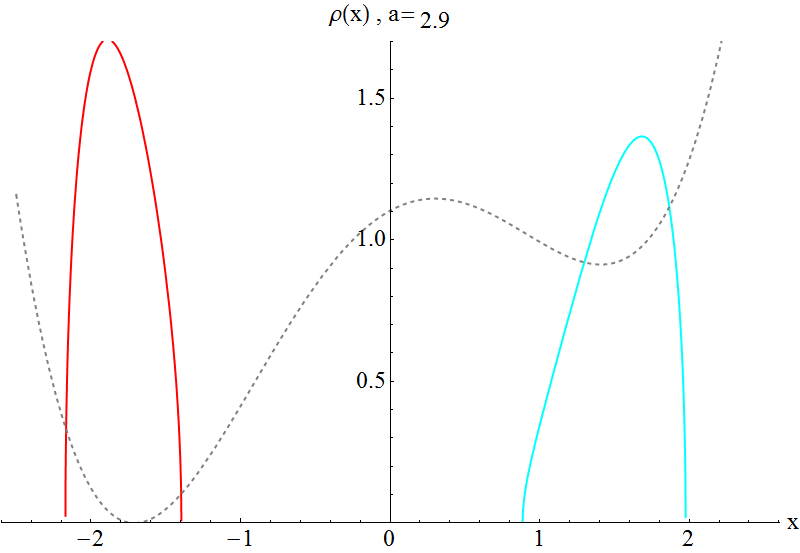}
\includegraphics[width=0.32\textwidth]{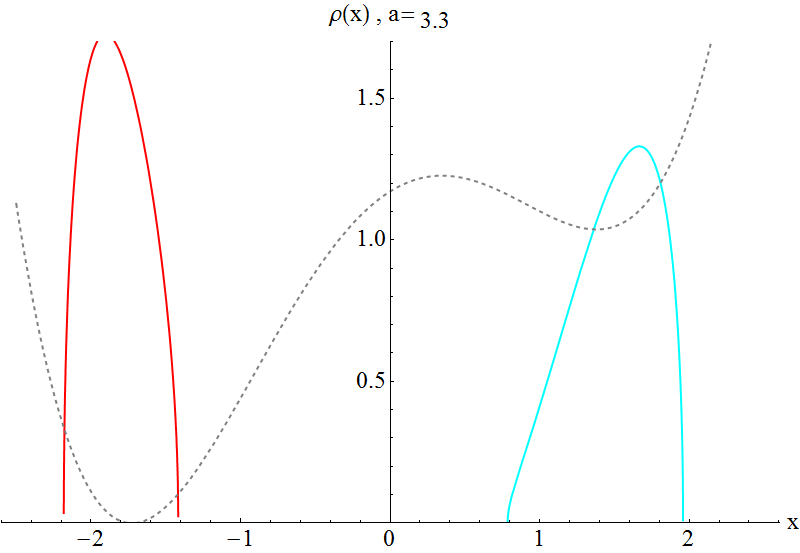}
\includegraphics[width=0.32\textwidth]{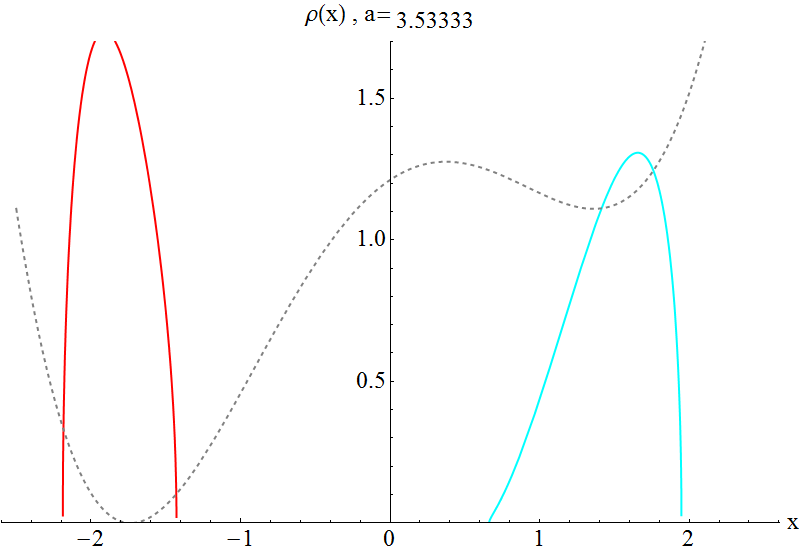}
\includegraphics[width=0.32\textwidth]{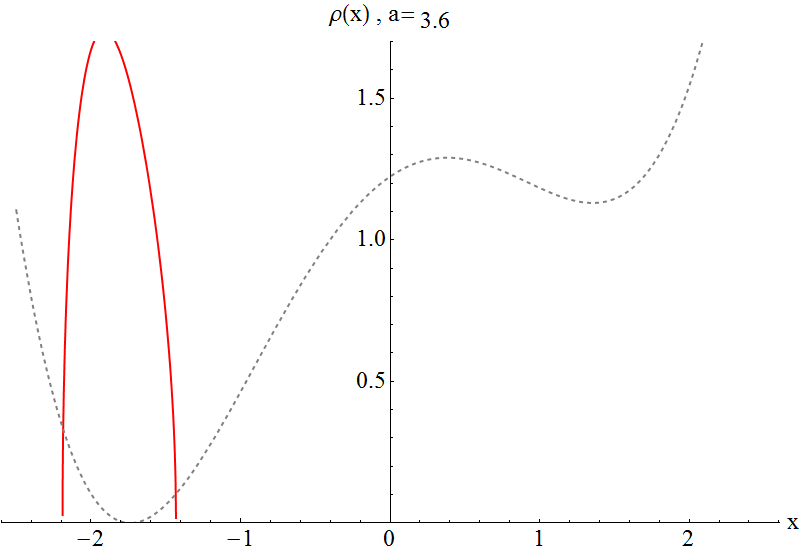}
\caption{Plots of eigenvalue distributions for the model (\ref{3actG1}) for values of ${r=-12}$ and various values of $a$, the dashed gray line denotes (a shifted and rescaled) potential. Note that all the two cut solutions interpolating between the left and the right one cut solutions also exist.}%
\label{fig_Bplots1}%
\end{figure}

\begin{figure}[H]%
\centering
\includegraphics[width=0.48\textwidth]{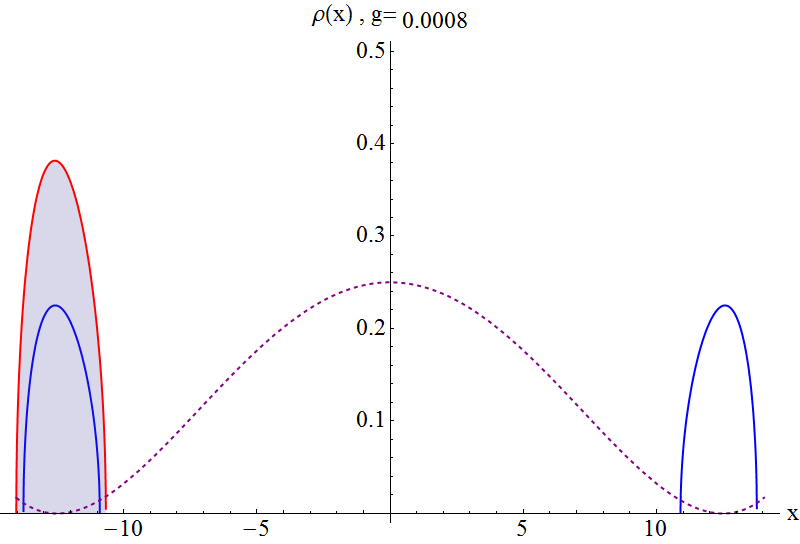}
\includegraphics[width=0.48\textwidth]{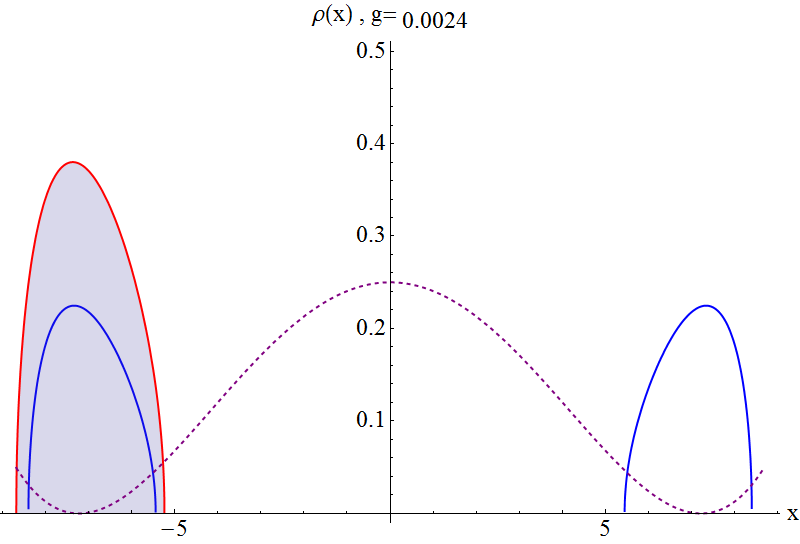}
\includegraphics[width=0.48\textwidth]{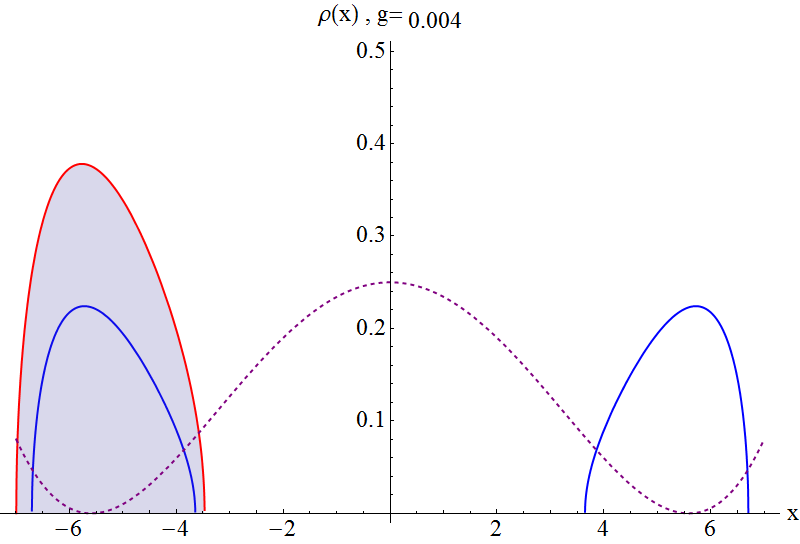}
\includegraphics[width=0.48\textwidth]{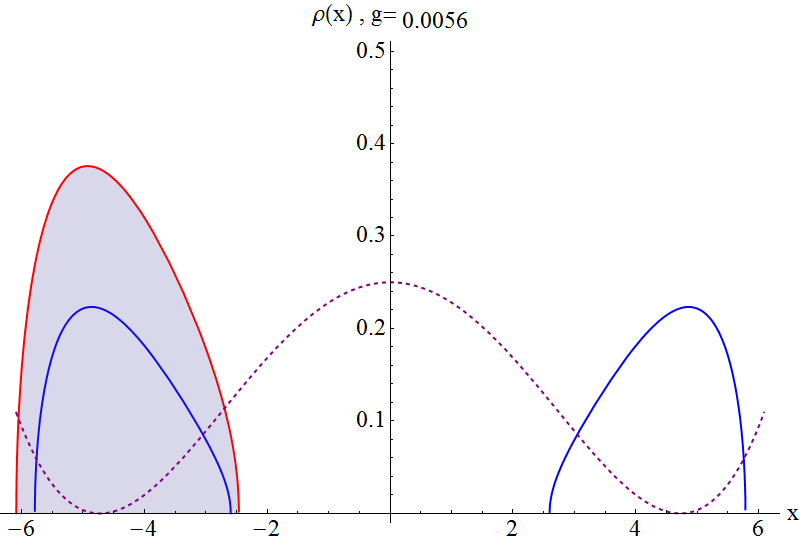}
\includegraphics[width=0.48\textwidth]{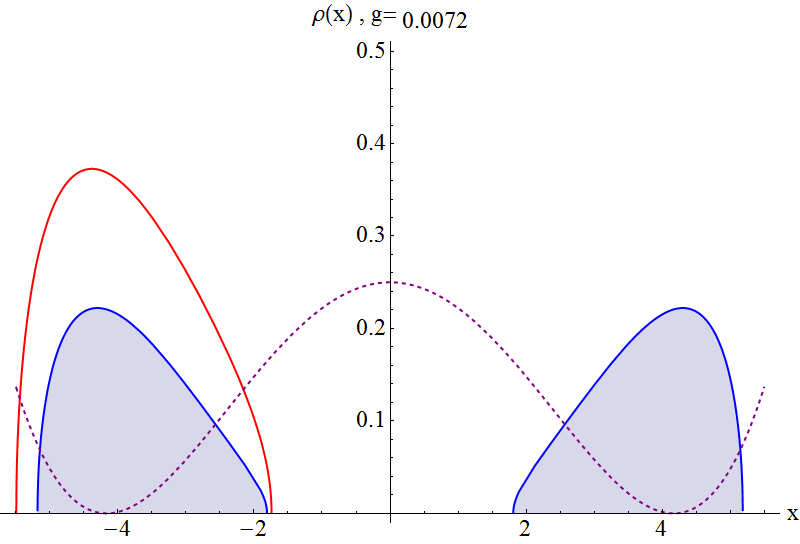}
\includegraphics[width=0.48\textwidth]{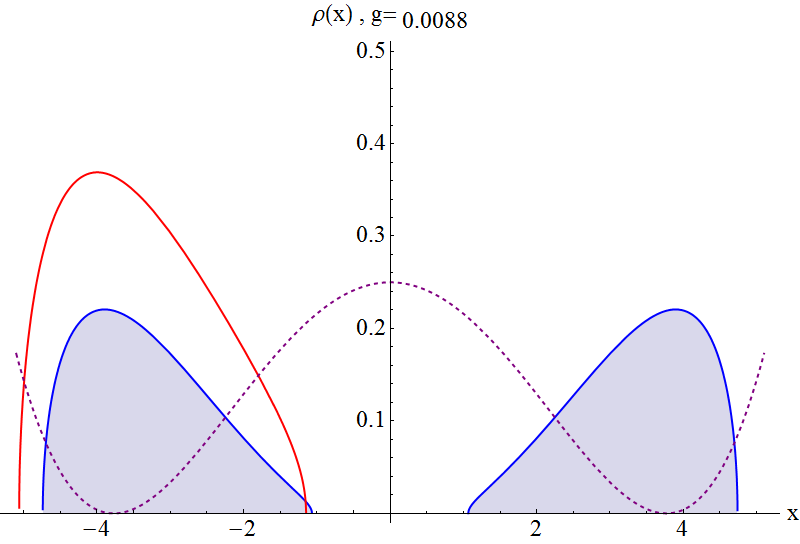}
\includegraphics[width=0.48\textwidth]{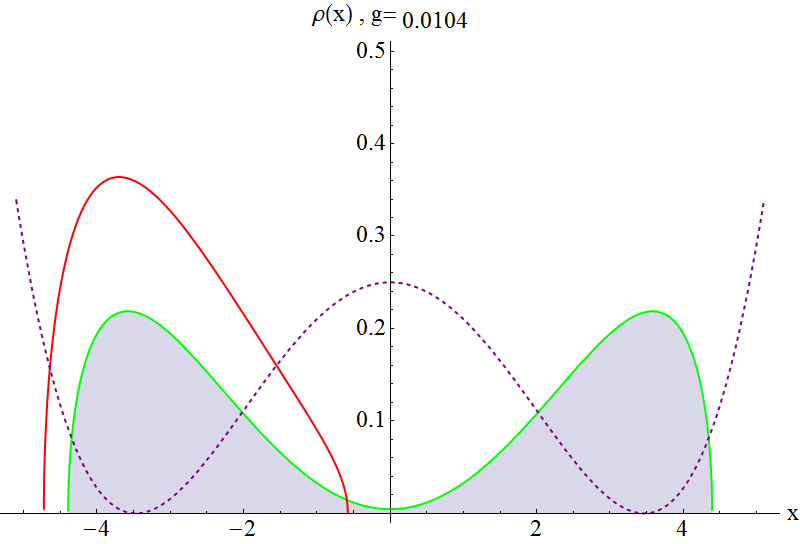}
\includegraphics[width=0.48\textwidth]{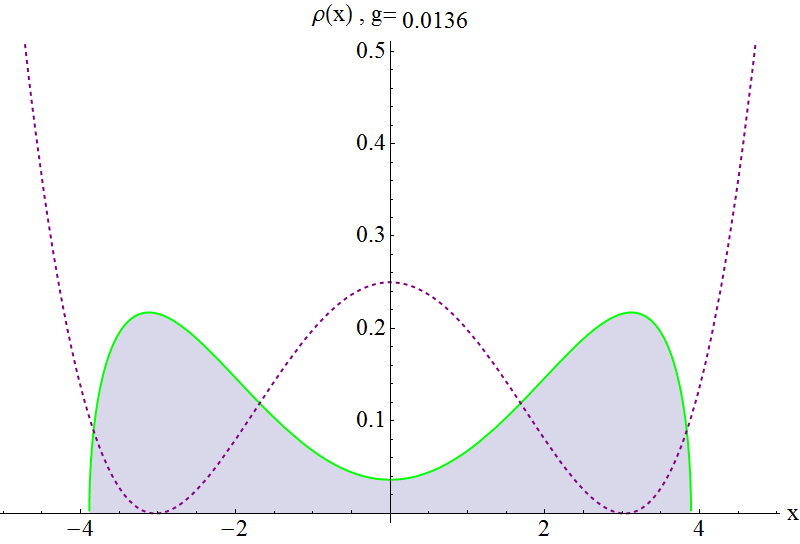}
\caption{Plots of eigenvalue distributions for the model (\ref{4model},\ref{5sphereF}) for values of ${r=-0.5}$ and various values of $g$, the dashed purple line denotes (a shifted and rescaled) potential. The filled solution denotes the preferred solution with the lower free energy.}%
\label{fig_Bplots2}%
\end{figure}


\end{document}